\documentclass[journal]{IEEEtran}
\usepackage{amsmath,amssymb,amsfonts}
\usepackage{algorithmic}
\usepackage{algorithm}
\usepackage{array}
\usepackage[caption=false,font=normalsize,labelfont=sf,textfont=sf]{subfig}
\usepackage{textcomp}
\usepackage{stfloats}
\usepackage{url}
\usepackage{verbatim}
\usepackage{comment}
\usepackage{graphicx}
\usepackage{cite}
\usepackage{enumitem}
\usepackage{booktabs}
\usepackage{multirow}
\usepackage{xcolor}
\newcommand{\TBD}[1]{#1}%
\newcommand{\scalefig}[2]{\includegraphics[width=#1\linewidth]{#2}}

\begin{document}

\title{Decoupled Coarse-to-Fine Learning for Modulo Signal Recovery}

\author{Yutao Jin,~\IEEEmembership{Graduate Student Member, IEEE}, Wenyi Yan,~\IEEEmembership{Student Member, IEEE}, Zichen Wang,~\IEEEmembership{Member, IEEE}, Hongqing Liu,~\IEEEmembership{Senior Member, IEEE}, Guoquan Li,~\IEEEmembership{Senior Member, IEEE}, and Lu Gan,~\IEEEmembership{Senior Member, IEEE}%
\thanks{This article has been accepted for publication in IEEE Transactions on Instrumentation and Measurement. \copyright~2026 IEEE. Personal use of this material is permitted. Permission from IEEE must be obtained for all other uses, in any current or future media, including reprinting/republishing this material for advertising or promotional purposes, creating new collective works, for resale or redistribution to servers or lists, or reuse of any copyrighted component of this work in other works.}%
\thanks{
Yutao Jin, Wenyi Yan, and Lu Gan are with the Department of Engineering, Brunel University of London, U.K. Zichen Wang is with the School of Electronic and Information Engineering, Tiangong University, China. Hongqing Liu and Guoquan Li are with the School of Communications and Information Engineering, Chongqing University of Posts and Telecommunications, China. }.
}

\markboth{}%
{Jin \MakeLowercase{\textit{et al.}}: Decoupled Coarse-to-Fine Learning for Modulo Signal Recovery}

\maketitle

\begin{abstract}
\TBD{Modulo analog-to-digital converters (ADCs) extend the dynamic range of measurement systems by folding out-of-range amplitudes into a bounded interval rather than clipping them. We propose CoFMo, a physics-inspired coarse-to-fine learning framework for high-fidelity modulo signal recovery at low oversampling factors. CoFMo jointly estimates absolute and relative fold counts before residual refinement, while a gated temporo-spectral module combines long-range temporal modeling with Wiener-inspired adaptive spectral processing. An uncertainty-based strategy further balances multiple recovery objectives without manual coefficient tuning. Extensive experiments show that CoFMo achieves the widest successful-recovery region among the evaluated analytical and learning-based baselines. On physical modulo ADC measurements with injected sensing noise and 4-bit quantization, CoFMo recovers 64-sample records with normalized mean-square errors of $-19.45$ to $-29.64$~dB, outperforming the strongest baseline by $7.81$--$10.89$~dB. To the best of our knowledge, this is the first learning-based framework validated on physical modulo ADC measurements, extending data-driven recovery from simulation to practical high-dynamic-range measurement.}
\end{abstract}


\section{Introduction}
\begin{figure}[t]
\centering
\def\figZeroScale{1.00} %
\def\figZeroLeftShift{-0.03\linewidth} %
\makebox[\linewidth][l]{\hspace*{\figZeroLeftShift}\includegraphics[width=\figZeroScale\linewidth]{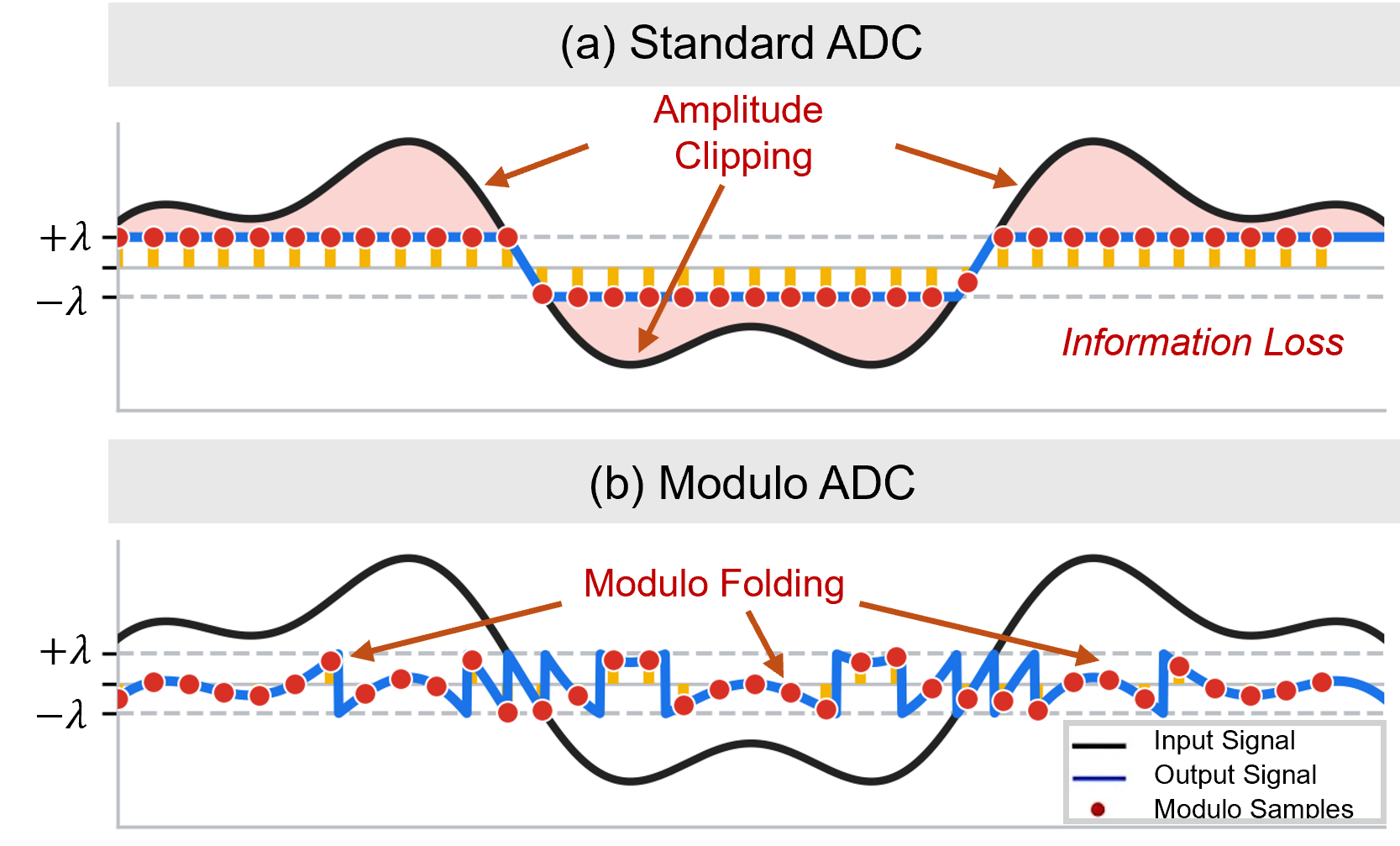}}
\caption{Conventional versus modulo sampling. (a) A conventional ADC irreversibly clips inputs beyond $\pm\lambda$. (b) A modulo ADC instead folds them into $[-\lambda, +\lambda)$ as modulo samples for later digital recovery.}
\label{fig0}
\end{figure}

\IEEEPARstart{M}{odern} measurement and sensing systems are required to faithfully acquire high-dynamic-range signals, ranging from broadband seismometers on NASA's InSight lander that record Marsquakes over wide amplitude spans \cite{lognonne2019seis, banerdt2020insight}, to accelerometers and ultrasonic transducers that encounter abrupt high-amplitude transients in vibration monitoring and non-destructive testing \cite{liang202516khz, liang202612bit}. Such acquisition relies on the analog-to-digital converter (ADC), which samples the input in time and quantizes it in amplitude over a finite dynamic range. Although the temporal resolution is characterized by the Shannon-Nyquist theorem \cite{shannon1949communication}, the amplitude dimension is often idealized as unbounded in theory. Practical ADCs, however, operate over a limited dynamic range, whose extension typically incurs trade-offs in sampling rate, power consumption, and circuit complexity \cite{walden1999,b1,gao202512}. As shown in Fig.~\ref{fig0}(a), when the input exceeds the ADC threshold $\lambda$, the measurement saturates or clips, causing irreversible loss of out-of-range information and ultimately degrading downstream sensing, estimation, and inference tasks \cite{ban2026error}.

To overcome this limitation, Bhandari \emph{et al.}~\cite{b5} proposed the Unlimited Sampling Framework (USF), also known as a modulo ADC, a hardware--algorithm co-design in which the front end applies a modulo operation prior to quantization instead of clipping, and the modulo signal is later recovered digitally. Mathematically, the front end folds the analog signal $x(t)$ with the centered modulo operator $\mathcal{M}_\lambda(\cdot)$,
\begin{equation}
y(t) = \mathcal{M}_{\lambda}\{x(t)\} = x(t) - 2\lambda \left\lfloor \frac{x(t)+\lambda}{2\lambda} \right\rfloor,
\label{eq1}
\end{equation}
where $\lfloor \cdot \rfloor$ is the floor function, and then samples the folded output at intervals $T_s$ to obtain the modulo samples $y[n] = y(nT_s)$. Whereas a conventional ADC clips out-of-range amplitudes, the modulo folding instead maps them back into the bounded interval $[-\lambda, +\lambda)$ and preserves them as modulo samples as depicted in Fig.~\ref{fig0}(b).

Recovering the unfolded signal from modulo samples requires resolving the ambiguity introduced by the modulo operation. The recovery difficulty is typically governed by the oversampling factor (OF) and the folding ratio $\rho$, which are defined as
\begin{equation}
\mathrm{OF} \triangleq \frac{f_s}{2\mathcal{B}},
\qquad
\rho \triangleq \frac{\|x\|_\infty}{\lambda},
\label{eq:OF_rho}
\end{equation}
where $f_s=1/T_s$ is the sampling frequency, $2\mathcal{B}$ is the Nyquist rate of a signal bandlimited $\mathcal{B}$, and $\|x\|_\infty$ denotes the peak amplitude of $x(t)$. The OF quantifies the temporal redundancy available for resolving the modulo ambiguity, whereas $\rho$ characterizes the severity of amplitude folding relative to the modulo threshold $\lambda$. Thus, modulo recovery is particularly challenging in the low-OF and large-$\rho$ regime, where severe folding must be resolved from limited temporal redundancy.

Existing recovery methods can be broadly divided into traditional analytical approaches and learning-based approaches. Analytical methods exploit temporal smoothness~\cite{b5, b2, romanov2019above} or spectral consistency~\cite{azar2022residual, shah2023lasso, zhang2024line, guo2023iter} to identify the fold counts, and they are effective when sufficient oversampling provides reliable local or global constraints. However, in the low-\(\mathrm{OF}\) and large-\(\rho\) regime, adjacent modulo samples may contain multiple or closely spaced folds, making the fold-count sequence difficult to disambiguate. \TBD{Moreover, practical modulo ADC measurements deviate from the ideal model because of front-end nonidealities, quantization and noise~\cite{HardwareFPGA, zhu2025modulo}}. These effects can accumulate in analytical pipelines and lead to unstable unfolding.

Learning-based methods provide an alternative route by learning the inverse mapping from modulo samples to unfolded signals directly from data~\cite{b11}. Nevertheless, they treat modulo recovery as a direct waveform regression problem and do not explicitly exploit the latent integer fold-count sequence. Moreover, they have been trained and evaluated only on synthetic modulo samples with added Gaussian noise and quantization, and their robustness to the nonidealities of practical modulo ADC measurements has not been demonstrated.

These limitations motivate a question: can physics-informed modelling be combined with data-driven learning to exploit the latent fold-count structure while remaining robust to practical modulo ADC nonidealities? To this end, this work investigates whether a physics-inspired learning framework can achieve robust modulo signal recovery from practical measurements at a low $\mathrm{OF}$. The main contributions are summarized as follows:

\begin{itemize}
  \item We propose CoFMo, a physics-inspired coarse-to-fine recovery framework that explicitly estimates the latent integer fold-count sequence before residual refinement, avoiding regression-based error propagation.

  \item Guided by the bandlimited signal model and the sparse fold-transition structure, we develop a gated temporo-spectral modelling strategy that combines dual-stream signal representations, adaptive spectral processing, and long-range temporal modelling for robust recovery.

  \item Extensive simulations and hardware measurements demonstrate state-of-the-art recovery performance over existing analytical and learning-based methods. To the best of our knowledge, this is the first learning-based framework validated on practical modulo ADC measurements, including stress tests under injected Gaussian noise and low-resolution quantization.
\end{itemize}

\textit{Notation.} Continuous signals are written as $x(t)$ and their uniformly sampled sequences as $x[n]$; $\lfloor\cdot\rfloor$ denotes the floor function, and $\|\cdot\|_\infty$ the supremum norm. The main symbols are summarized in Table~\ref{tab:notation}.

\begin{table}[!t]
\renewcommand{\arraystretch}{1.3}
\caption{Summary of Notation}
\label{tab:notation}
\centering
\begin{tabular}{l|p{6cm}}
\toprule
\hline
\textbf{Notation} & \textbf{Description} \\
\hline
$\varepsilon[n]$ & Additive measurement perturbation \\
\hline
$x[n]\,/\,\hat{x}[n]$ & Clean\,/\,recovered signal \\
\hline
$\lambda$ & Modulo folding threshold \\
\hline
$\mathcal{M}_\lambda(\cdot)$ & Centered modulo operator, $\mathcal{M}_\lambda(x)=x-2\lambda k$ \\
\hline
$y[n]\,/\,y_{\mathrm{e}}[n]$ & Clean\,/\,noisy modulo samples \\
\hline
$k[n]\,/\,\hat{k}[n]$ & True\,/\,estimated integer fold count \\
\hline
$\hat{r}[n]$ & Estimated residual correction \\
\hline
$x_{\mathrm{coarse}}[n]$ & Coarse unwrapped estimate, $y[n]+2\lambda\hat{k}[n]$ \\
\hline
$\Delta^N$ & $N$-th order finite difference \\
\hline
$K_{\max}\,/\,\Delta k_{\max}$ & Absolute\,/\,relative fold budget\\
\hline
$\mathcal{T}_{\Theta}(\cdot)$ & Adaptive spectral gating (shrinkage) operator \\
\hline
$s_i$ & Per-task uncertainty weight, $s_i=\log\sigma_i^2$ \\
\hline
$L$ & Sequence length (samples per signal) \\
\hline
$\mathcal{B}$ & Signal bandwidth \\
\hline
 $T_s$ & Sampling interval \\
\hline
$f_s$ & Sampling frequency, $f_s = 1/T_s$ \\
\hline
$\rho$ & Folding ratio, $\rho=\|x\|_\infty/\lambda$ \\
\hline
$\mathrm{OF}$ & Oversampling factor, $\mathrm{OF}=f_s/(2\mathcal{B})$ \\
\hline
$\mathrm{SNR}_x$ & Signal-to-noise ratio w.r.t.\ the original signal $x[n]$ \\
\hline
$\mathrm{SNR}_y$ & Signal-to-noise ratio w.r.t.\ the modulo signal $y[n]$ \\
\hline
\bottomrule
\end{tabular}
\end{table}

\section{Related Work}
\label{sec:related}

This section reviews traditional analytical approaches and recent learning-based algorithms for modulo signal recovery.

\subsection{Analytical Modulo Signal Recovery}
\label{sec:analytical_recovery}

Analytical methods exploit the structure of the folding operation. The clean modulo decomposition is given by:
\begin{equation}
y[n]=\mathcal{M}_{\lambda}\!\left(x[n]\right)=x[n]-2\lambda k[n], \qquad k[n]\in\mathbb{Z}.
\label{eq:decomp}
\end{equation}
To account for practical nonidealities, the measured modulo samples can be modeled as
\begin{equation}
y_{\mathrm{e}}[n] \triangleq y[n]+\varepsilon[n]
= x[n]-2\lambda k[n]+\varepsilon[n],
\label{eq:noisy}
\end{equation}
where $\varepsilon[n]$ includes analogue distortion, timing jitter, measurement noise, and quantisation error.

Let $\hat{k}[n]$ denote the fold-count estimate obtained from $y_{\mathrm{e}}[n]$. The corresponding unfolded estimate is
\begin{equation}
\hat{x}[n]=y_{\mathrm{e}}[n]+2\lambda \hat{k}[n].
\label{eq:estimated_unfolding}
\end{equation}
Any fold-count error $\hat{k}[n]\neq k[n]$ induces a reconstruction error equal to an integer multiple of $2\lambda$, whereas $\varepsilon[n]$ contributes only a bounded residual; accurate fold-count recovery thus governs the overall reconstruction fidelity.

\subsubsection{Near-Nyquist Identifiability and Prediction-Based Recovery}

A bandlimited signal is identifiable from its modulo samples at rates arbitrarily close to Nyquist~\cite{bhandari2019identifiability}. Romanov and Ordentlich~\cite{romanov2019above} proposed a prediction-based scheme above Nyquist that initialises a causal predictor from an initial unfolded segment (samples known to lie within the ADC threshold). It substantially lowers the noiseless sampling rate, but its unbounded predictor gain renders the recovery numerically unstable, amplifying small perturbations into large errors and leaving it sensitive to noise and hardware non-idealities.

\subsubsection{Higher-Order-Difference Recovery}

The original Unlimited Sampling algorithm exploits the commutativity between integer-coefficient finite differences and the modulo operation~\cite{b5}. With $\Delta$ the first-order difference, $(\Delta x)[n]=x[n]-x[n-1]$, and $\Delta^N$ its $N$-fold application, the clean model in~\eqref{eq:decomp} gives
\begin{equation}
\begin{aligned}
\Delta^N y[n]
&= \Delta^N x[n]-2\lambda\Delta^N k[n], \\
\mathcal{M}_{\lambda}\!\left(\Delta^N y[n]\right)
&= \Delta^N x[n],
\qquad \text{if } \|\Delta^N x\|_{\infty}<\lambda .
\end{aligned}
\label{eq:hod_review}
\end{equation}
Once $N$ is large enough to keep the signal term within one modulo interval, the fold ambiguity is removed and $x[n]$ recovered by recursive summation.

The bound $\|\Delta^N x\|_{\infty}\leq (\pi e/\mathrm{OF})^N\|x\|_{\infty}$~\cite{b5} gives the order-dependent condition $\rho<(\mathrm{OF}/\pi e)^N$, so US-ALG requires substantial oversampling, $\mathrm{OF}\approx2\pi e$. A tightened analysis with a revised second-order difference~\cite{b2} lowers the noiseless requirement to $\mathrm{OF}>3$.

Robustness, however, remains limited. A larger $\rho$ demands a higher order $N$, yet each difference amplifies noise by up to $2^N$. Once the signal exceeds the order-dependent threshold, fold-estimation errors propagate through the recursive summation and cause catastrophic reconstruction~\cite{b5,b2}.

\subsubsection{First-Order-Difference Recovery and Signal Priors}

To avoid high-order differencing, other methods exploit the first-order difference. Applying $\Delta$ to the clean and perturbed models~\eqref{eq:decomp},\,\eqref{eq:noisy} yields
\begin{equation}
\begin{aligned}
\Delta y[n]
&= \Delta x[n]-2\lambda\Delta k[n], \\
\Delta y_{\mathrm{e}}[n]
&= \Delta x[n]-2\lambda\Delta k[n]+\Delta\varepsilon[n],
\end{aligned}
\label{eq:fold_diff}
\end{equation}
where $\Delta k[n]=k[n]-k[n-1]$. This exposes complementary priors underlying most first-order methods. i) The increment $\Delta x[n]$ stays \textit{bandlimited} to the original signal band, though finite differencing emphasises higher in-band frequencies. ii) For moderate folding ratios ($\rho\lesssim 20$ here) and dense sampling, folds occur only at boundary crossings, so the fold-transition sequence $\Delta k[n]$ stays \textit{sparse} with most entries zero.

Threshold-based approaches directly exploit these properties~\cite{florescu2022thresholding, florescu2022hysteresis, rudresh2018wavelet}. Generalized thresholding detects folds from the structured transitions of a filtered modulo signal, the hysteretic modulo model enforces a minimum separation between folds to improve detectability on non-ideal hardware, and wavelet-domain recovery exploits the increment's smoothness to suppress folding-induced discontinuities.

Spectral approaches instead estimate the sparse fold-transition sequence from the out-of-band residual. The B$^2$R$^2$ principle exploits the out-of-band spectrum of the differenced samples~\cite{azar2022residual}, on which LASSO-B2R2 applies sparse regularisation~\cite{shah2023lasso,shah2024compressed}, Fourier--Prony parametric spectral modelling~\cite{bhandari2021unlimited}, USLSE integer least squares for line spectra~\cite{zhang2024line}, and ITER-SIS iterative time-domain sieving~\cite{guo2023iter}.

All of these require a clear separation between the smooth signal and the sparse fold transitions, which weakens at a low $\mathrm{OF}$ as temporal and spectral redundancy shrinks. Practical measurements also carry the term $\Delta\varepsilon[n]$ in~\eqref{eq:fold_diff}. Being high-pass, differencing amplifies it and renders fixed thresholds, sparsity assumptions, and model-based tuning unreliable.

\subsubsection{Implications for Practical Recovery}

These properties carry over to a data-driven design. The reconstruction is governed by an underlying integer fold-count sequence, and the pair $(y_{\mathrm{e}}[n],\Delta y_{\mathrm{e}}[n])$ supplies complementary amplitude and fold-transition cues from which the clean fold structure $(k[n],\Delta k[n])$ must be recovered, even though the measurements are corrupted by the perturbation $\varepsilon[n]$. This motivates a physics-informed design that preserves these analytical priors while remaining robust at a low $\mathrm{OF}$ on measurements.

\subsection{Learning-Based Modulo Signal Recovery}

\TBD{Learning-based methods have been widely studied for high-dynamic-range (HDR) signal and image processing \cite{mai2022deep, zhou2026high, chung2023lan, robidoux2021end, wang2024ufc}. More closely related to modulo recovery, UnModNet \cite{b8}, the gradient-domain unfolding method in \cite{chen2025robust}, and the lightweight unrolled network in \cite{monroy2026deep} exploit rollover masks, gradient continuity, or finite-difference priors for modulo imaging. These methods are primarily designed for two-dimensional spatial signals and rely on image-domain structures that do not directly apply to one-dimensional bandlimited modulo samples acquired at a low $\mathrm{OF}$.}

\TBD{Learning-based recovery for 1D modulo sampling remains much less explored. Koenen et al. \cite{b11} proposed MUNet, a convolutional residual network that estimates the difference between the original and modulo signals. MUNet mainly formulates recovery as continuous signal regression, incorporating the modulo structure through a sparsity penalty on the first-order difference of the predicted residual and post-hoc rounding to $2\lambda\mathbb{Z}$. It was evaluated on synthetic modulo data without validation on physical modulo ADC measurements.}

\TBD{CoFMo differs from existing learning-based approaches in four aspects. First, it incorporates the modulo physics explicitly through both modulo amplitudes and first-order differences. Second, the absolute fold counts and their temporal differences are modeled as discrete classification variables rather than inferred indirectly from continuous regression. Third, the gated temporo-spectral module captures long-range sequence dependencies while exploiting spectral structure for robust recovery at low $\mathrm{OF}$. Finally, CoFMo is trained and validated on physical modulo ADC measurements, including hardware nonidealities, injected sensing noise, and low-resolution quantization.}

\section{Method}
\label{sec:method}

Recovering the signal $\hat{x}[n]$ from the noisy modulo samples $y_{\mathrm{e}}[n]$ is equivalent to resolving the estimated fold-count sequence $\hat{k}[n]$. Because amplitudes differing by integer multiples of $2\lambda$ produce identical observations, this inversion is a many-to-one mapping that becomes ill-posed and susceptible to folding errors when $\hat{k}[n]$ is inferred from isolated samples. To address this, CoFMo first performs a discrete classification that predicts the fold counts for a coarse unfolding, and then a continuous regression that refines the sub-fold residual.

\begin{figure*}[htbp]
\centering
\def\figTwoScale{0.9} %
\includegraphics[width=\figTwoScale\linewidth]{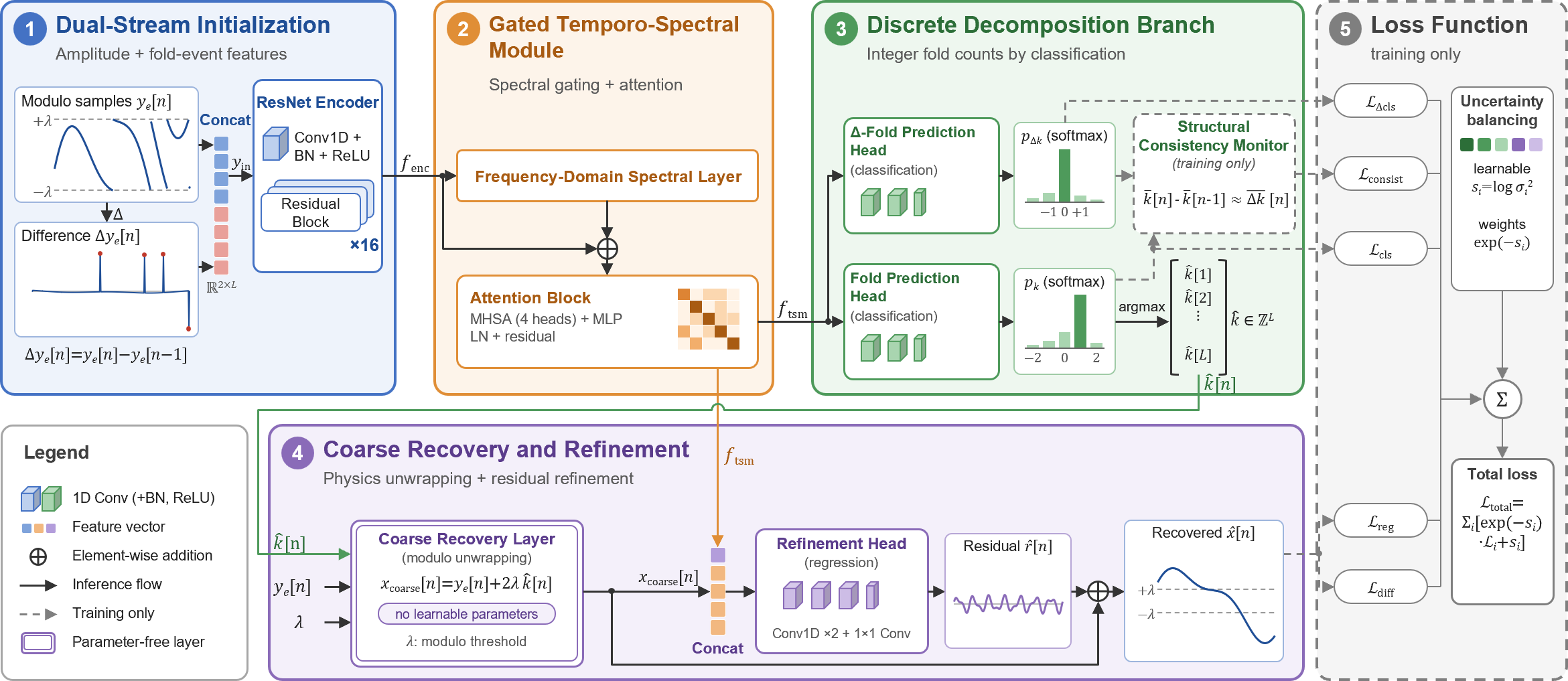}
\caption{\TBD{Workflow of the proposed CoFMo framework. \emph{Part 1 (Dual-Stream Initialization):} the noisy modulo signal $y_{\mathrm{e}}[n]$ and its first-order difference are encoded by a ResNet into the initial features $f_{\mathrm{enc}}$. \emph{Part 2 (Gated Temporo-Spectral Module):} $f_{\mathrm{enc}}$ is processed by an adaptive spectral gate and multi-head self-attention to yield the contextual features $f_{\mathrm{tsm}}$. \emph{Part 3 (Discrete Decomposition Branch):} dual classification heads predict the absolute ($\hat{k}$) and relative ($\Delta\hat{k}$) fold counts, coupled by a structural-consistency constraint. \emph{Part 4 (Coarse Recovery and Refinement):} $\hat{k}$ gives the closed-form coarse estimate $x_{\mathrm{coarse}}$, which is concatenated with $f_{\mathrm{tsm}}$ and refined by a regression head into the residual $r$, yielding the recovered signal $\hat{x}[n]$. The five loss terms are balanced by an uncertainty-driven weighting scheme with learnable parameters $\log\sigma_i^2$.}}
\label{fig2}
\end{figure*}

\subsection{CoFMo Architecture}
As shown in Fig. \ref{fig2}, the proposed CoFMo framework is a hierarchical architecture. It comprises a ResNet encoder, a gated temporo-spectral module, a parallel discrete decomposition branch, and a coarse-to-fine recovery pipeline, collectively optimized via an uncertainty-balanced multi-task objective. Given the wrapped noisy observation $y_{\mathrm{e}} = \{y_{\mathrm{e}}[n]\}_{n=1}^{L}$, where $L$ is the sequence length, we seek to recover the original signal $x = \{x[n]\}_{n=1}^{L}$. We formulate this as jointly estimating an absolute fold-count sequence $\hat{k} = \{\hat{k}[n]\}_{n=1}^{L}$, a relative fold-change sequence $\Delta\hat{k} = \{\Delta\hat{k}[n]\}_{n=2}^{L}$, and an estimated residual $\hat{r} = \{\hat{r}[n]\}_{n=1}^{L}$ that corrects for sub-fold reconstruction errors. The overall forward pass proceeds as $y_{\mathrm{e}}[n] \to f_{\mathrm{enc}} \to f_{\mathrm{tsm}} \to (\hat{k}, \Delta\hat{k}, x_{\mathrm{coarse}}) \to \hat{x}[n]$, where each intermediate representation serves a distinct physical and statistical role in the unwrapping process. The final recovered signal can be expressed as:
\begin{equation}
    \hat{x}[n]  =  y_{\mathrm{e}}[n]  +  2\lambda \cdot \hat{k}[n]  +  \hat{r}[n]
\end{equation}
where $\hat{k}[n]$ is the predicted fold count and $\hat{r}[n]$ is the predicted residual. The sum $y_{\mathrm{e}}[n] + 2\lambda \cdot \hat{k}[n]$ forms the coarse unwrapped estimate $x_{\mathrm{coarse}}[n]$ for subsequent regression refinement.

We next detail the dual-stream initialization, the gated temporo-spectral module, the discrete decomposition branch, the coarse recovery and refinement, and the loss functions.

\subsection{Dual-Stream Initialization}
Learning directly from the noisy folded sequence $y_{\mathrm{e}}[n]$ is difficult because of the abrupt discontinuities at the folding boundaries. To emphasize these transitions, we adopt the first-order difference $\Delta y_{\mathrm{e}}[n]=y_{\mathrm{e}}[n]-y_{\mathrm{e}}[n-1]$, which by Eq.~\eqref{eq:fold_diff} isolates the sparse fold term $2\lambda\,\Delta k[n]$, nonzero only at the folding instants, from the small and slowly varying $\Delta x[n]$. The difference stream thus acts as a fold-event detector and exposes the discrete fold-change $\Delta k[n]$ estimated by the relative-fold head, a structure widely used in analytical methods.

In practice, $y_{\mathrm{e}}[n]$ is corrupted by the measurement noise and hardware non-idealities. As a high-pass operator, the first-order difference sharpens the folding edges but at the same time amplifies these perturbations, so the sparse fold cues are partly masked and a fixed threshold on $\Delta y_{\mathrm{e}}[n]$ becomes unreliable at the low oversampling factors. Therefore, we retain both streams rather than either alone. The raw stream $y_{\mathrm{e}}[n]$ provides a noise-robust amplitude trajectory, while the difference stream $\Delta y_{\mathrm{e}}[n]$ localizes the folding events. The two streams are concatenated along the channel dimension as:
\begin{equation}
    y_{\mathrm{in}}[n]=\mathrm{Concat}\big(y_{\mathrm{e}}[n],\,\Delta y_{\mathrm{e}}[n]\big),
\end{equation}
where $\Delta y_{\mathrm{e}}[n]$ is zero-padded at the boundary. This yields a tensor $f_{\mathrm{in}}\in\mathbb{R}^{B\times 2\times L}$ for a batch size $B$ and sequence length $L$, which an initial 1-D convolution with batch normalization maps to a $C$-channel embedding $f_{c}\in\mathbb{R}^{B\times C\times L}$, with $C=64$.

Then, the embedding $f_c$ is processed by a $16$-block residual encoder, which acts as a learnable adaptive filter that suppresses the residual perturbations while preserving the fold-induced transitions. With $\mathbf{h}_0=f_c$, the $\ell$-th block computes
\begin{equation}
    \mathbf{h}_\ell=\sigma\big(\mathbf{h}_{\ell-1}+\mathcal{R}_\ell(\mathbf{h}_{\ell-1})\big),
\end{equation}
\begin{equation}
    \mathcal{R}_\ell(\mathbf{h}_{\ell-1})
    =\mathrm{BN}_2\!\big(\mathbf{W}^{(\ell)}_2 * \sigma(\mathrm{BN}_1(\mathbf{W}^{(\ell)}_1 * \mathbf{h}_{\ell-1}))\big),
\end{equation}
where $\sigma$ is the ReLU activation, $*$ is the 1-D convolution with kernel size $K=7$, zero-padded to preserve the length $L$, $\mathbf{W}^{(\ell)}_1$ and $\mathbf{W}^{(\ell)}_2$ are the convolutional weights for the $\ell$-th block, and $\mathrm{BN}_1$ and $\mathrm{BN}_2$ represent Batch Normalization layers, respectively. The skip connection lets each block apply an incremental correction instead of re-estimating the whole feature map, which eases optimization and lets the encoder progressively suppress high-frequency distortion while retaining the folding cues. The output of the last block, $f_{\mathrm{enc}}=\mathbf{h}_{16}$, is passed to the temporo-spectral module.

\begin{figure}[t]
\centering
\scalefig{0.8}{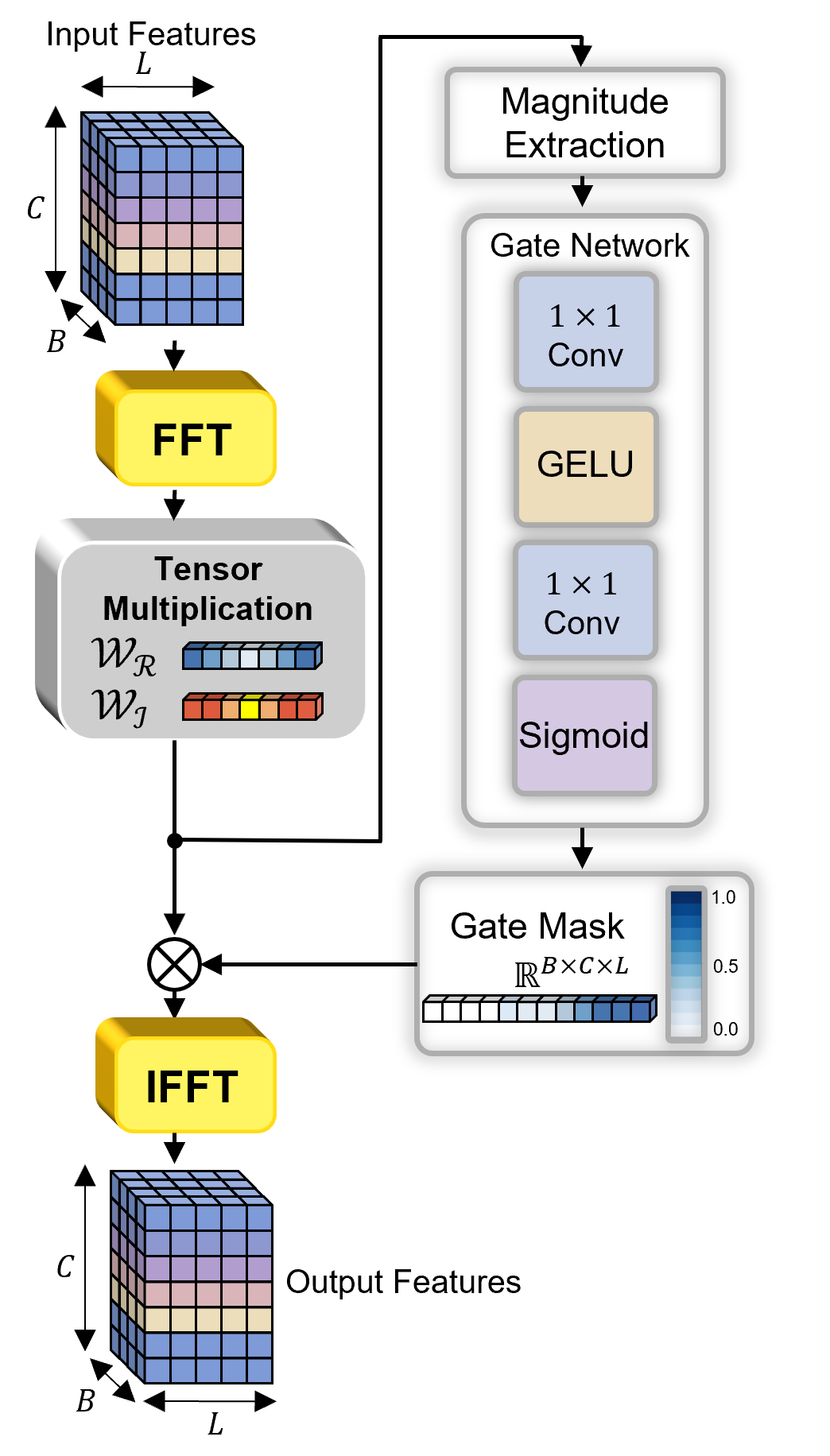}
\caption{Architecture of the Frequency-domain Spectral layer. Time-domain input features are mapped via FFT and multiplied with complex weights. A gate network then processes the magnitude response to generate an adaptive mask, which modulates the complex spectrum before IFFT reconstruction.}
\label{fig2_2}
\end{figure}

\subsection{Gated Temporo-Spectral Module}
Modulo folding induces structured signatures in both the time and
frequency domains. The folding events appear as sparse, abrupt transitions
whose energy extends beyond the signal bandwidth, whereas the fold-count
sequence $k[n]$ is a non-local quantity governed by the accumulated folding
history of the waveform. As no single-domain operator can resolve both
structures, we couple adaptive spectral filtering with temporal
self-attention within a gated temporo-spectral module. The spectral branch
is applied to the learned representation $f_{\mathrm{enc}}$ rather than to
the raw measurement, since the bandlimited prior underlying analytical
recovery is valid only under idealized conditions, whereas in practice
hardware non-idealities such as analog nonlinearity, quantization, and
folding transients introduce broadband perturbations that render a direct
Fourier transform of $y_{\mathrm{e}}[n]$ unreliable. By acting as a multi-channel filter
bank that distributes the folding and noise components across its $C$
channels, the encoder frames the spectral processing of $f_{\mathrm{enc}}$ as a learned generalization of analytical frequency-domain recovery.

As illustrated in Fig.~\ref{fig2_2}, we map $f_{\mathrm{enc}}$ to the
spectral domain through the discrete Fourier transform (DFT), which
reformulates the discrete sequence into a complex frequency representation
$F_{\mathrm{enc}} \in \mathbb{C}^{B \times C \times L}$ over the frequency
bins $v \in \{0, 1, \dots, L-1\}$,
\begin{equation}
    F_{\mathrm{enc}}[v] = \sum_{n=0}^{L-1} f_{\mathrm{enc}}[n] \exp\left( -j \frac{2\pi}{L} v n \right).
\end{equation}
To aggregate global context, we perform a tensor multiplication directly
in the frequency domain using complex-valued learnable weights and biases,
parameterized by their real and imaginary components $\mathcal{W}_{\mathcal{R}}$
and $\mathcal{W}_{\mathcal{J}}$. For each frequency bin $v$, this projects
the $C$ input channels into an intermediate complex state
$\mathbf{U} \in \mathbb{C}^{B \times C \times L}$; together with the
frequency-dependent complex bias $\mathcal{B}[v]$, the dense interaction reads
\begin{equation}
\begin{aligned}
\mathbf{U}[v] = \mathcal{B}[v]
&+ \sum_{c=1}^{C} \Bigg(
\mathrm{Re}\big(F_{\mathrm{enc}, c}[v]\big) \circledast \mathcal{W}_{\mathcal{R}, c}[v] \\
&\qquad\quad - \mathrm{Im}\big(F_{\mathrm{enc}, c}[v]\big) \circledast \mathcal{W}_{\mathcal{J}, c}[v] \\
&\qquad\quad + j \Big(
\mathrm{Re}\big(F_{\mathrm{enc}, c}[v]\big) \circledast \mathcal{W}_{\mathcal{J}, c}[v] \\
&\qquad\qquad\; + \mathrm{Im}\big(F_{\mathrm{enc}, c}[v]\big) \circledast \mathcal{W}_{\mathcal{R}, c}[v]
\Big) \Bigg),
\end{aligned}
\end{equation}
where $\circledast$ denotes the tensor contraction along the channel
dimension. While this dense interaction aggregates global context, it can
also amplify unreliable responses, since under low SNR the noise energy is
spread across the spectrum and corrupts the learned features.

Motivated by the Wiener filtering \cite{wiener1949extrapolation, scalart1996speech}, we formulate a gate in a Wiener-like form while learning the associated pseudo-SNR end-to-end from the spectral features. We first extract the element-wise magnitude of $\mathbf{U}$,
yielding the amplitude spectrum
$\mathcal{A} \in \mathbb{R}^{B \times C \times L}$ that serves as a proxy
for the frequency-bin energy,
\begin{equation}
\mathcal{A} = \sqrt{ \mathrm{Re}(\mathbf{U})^2 + \mathrm{Im}(\mathbf{U})^2 }.
\end{equation}

In a classical frequency-domain Wiener estimator, the gain is
$G=\xi/(\xi+1)$, where $\xi$ denotes the \textit{a priori}
SNR \cite{scalart1996speech}. We retain this bounded, monotonic gain
structure but replace the model-based SNR with a task-adaptive pseudo-SNR
learned from the feature-magnitude profile. To achieve it, we estimate a data-driven log-SNR from
$\mathcal{A}$ through a lightweight gate network composed of a $1\times1$
convolution, a GELU activation, and a second $1\times1$ convolution,
denoted $\mathbf{\Phi}_{\Theta}(\mathcal{A}) \in \mathbb{R}^{B \times C \times L}$,
\begin{equation}
    \mathbf{\Phi}_{\Theta}(\mathcal{A}) = \mathbf{W}_{1\times1}^{(2)} \ast \delta_{\mathrm{GELU}}\bigg( \mathbf{W}_{1\times1}^{(1)} \ast \mathcal{A} + \mathbf{b}^{(1)} \bigg) + \mathbf{b}^{(2)},
\end{equation}
where $\mathbf{W}_{1\times1}^{(1)}$ and $\mathbf{W}_{1\times1}^{(2)}$ are
the convolution kernels. Defining the data-driven pseudo-SNR tensor
$\mathbf{\Xi} = \exp(\mathbf{\Phi}_{\Theta}(\mathcal{A}))$, the proposed
Wiener-like gain is given by
\begin{equation}
\mathbf{G} = \sigma_{\mathrm{Sig}}\big(\mathbf{\Phi}_{\Theta}(\mathcal{A})\big) = \frac{\exp(\mathbf{\Phi}_{\Theta}(\mathcal{A}))}{\exp(\mathbf{\Phi}_{\Theta}(\mathcal{A})) + \mathbf{1}} \equiv \frac{\mathbf{\Xi}}{\mathbf{\Xi} + \mathbf{1}}.
\end{equation}
The refined spectral representation $\tilde{\mathbf{U}}$ then follows from
element-wise modulation, which we cast as a parametric shrinkage operator
$\mathcal{T}_{\Theta}(\mathbf{U})$ \cite{donoho1995noising},
\begin{equation}
\tilde{\mathbf{U}} = \mathcal{T}_{\Theta}(\mathbf{U}) := \mathbf{G} \odot \mathbf{U} = \left( \frac{\mathbf{\Xi}}{\mathbf{\Xi} + \mathbf{1}} \right) \mathbf{U}.
\end{equation}

Acting as a spectral estimator driven by  $\mathcal{A}$,
this operator selectively attenuates unreliable components while preserving
informative structures. Its behavior becomes transparent through an
element-wise asymptotic analysis. For any element $i$ with local
pseudo-SNR $\mathbf{\Xi}_i$, the gated response satisfies
\begin{equation}
[\tilde{\mathbf{U}}]_i =
\begin{cases}
\mathbf{\Xi}_i [\mathbf{U}]_i + \mathcal{O}\left(\mathbf{\Xi}_i^2\right), & \text{if } \mathbf{\Xi}_i \ll 1 \\
[\mathbf{U}]_i - \mathbf{\Xi}_i^{-1} [\mathbf{U}]_i + \mathcal{O}\left(\mathbf{\Xi}_i^{-2}\right), & \text{if } \mathbf{\Xi}_i \gg 1
\end{cases}
\end{equation}
In noise-dominated regions ($\mathbf{\Xi}_i \ll 1$), the operator acts as
an aggressive linear attenuator proportional to the pseudo-SNR, whereas in
signal-dominated regions ($\mathbf{\Xi}_i \gg 1$) it approaches an identity
mapping with a suppression error that decays inversely with the
pseudo-SNR. The mechanism thus performs an implicit, noise-aware shrinkage
in the spectral domain, retaining salient signal harmonics while
discarding noise-dominated components. Finally, the modulated features are
returned to the time domain through the inverse DFT, yielding
$f_{\mathrm{out}} \in \mathbb{R}^{B \times C \times L}$,
\begin{equation}
\begin{aligned}
    f_{\mathrm{out}}[n]
    &= \frac{1}{L} \sum_{v=0}^{L-1} \tilde{\mathbf{U}}[v] \exp\left( j \frac{2\pi}{L} v n \right), \\
    &\qquad \forall n \in \{0, 1, \dots, L-1\}.
\end{aligned}
\end{equation}

\begin{figure}[t]
\centering
\scalefig{0.75}{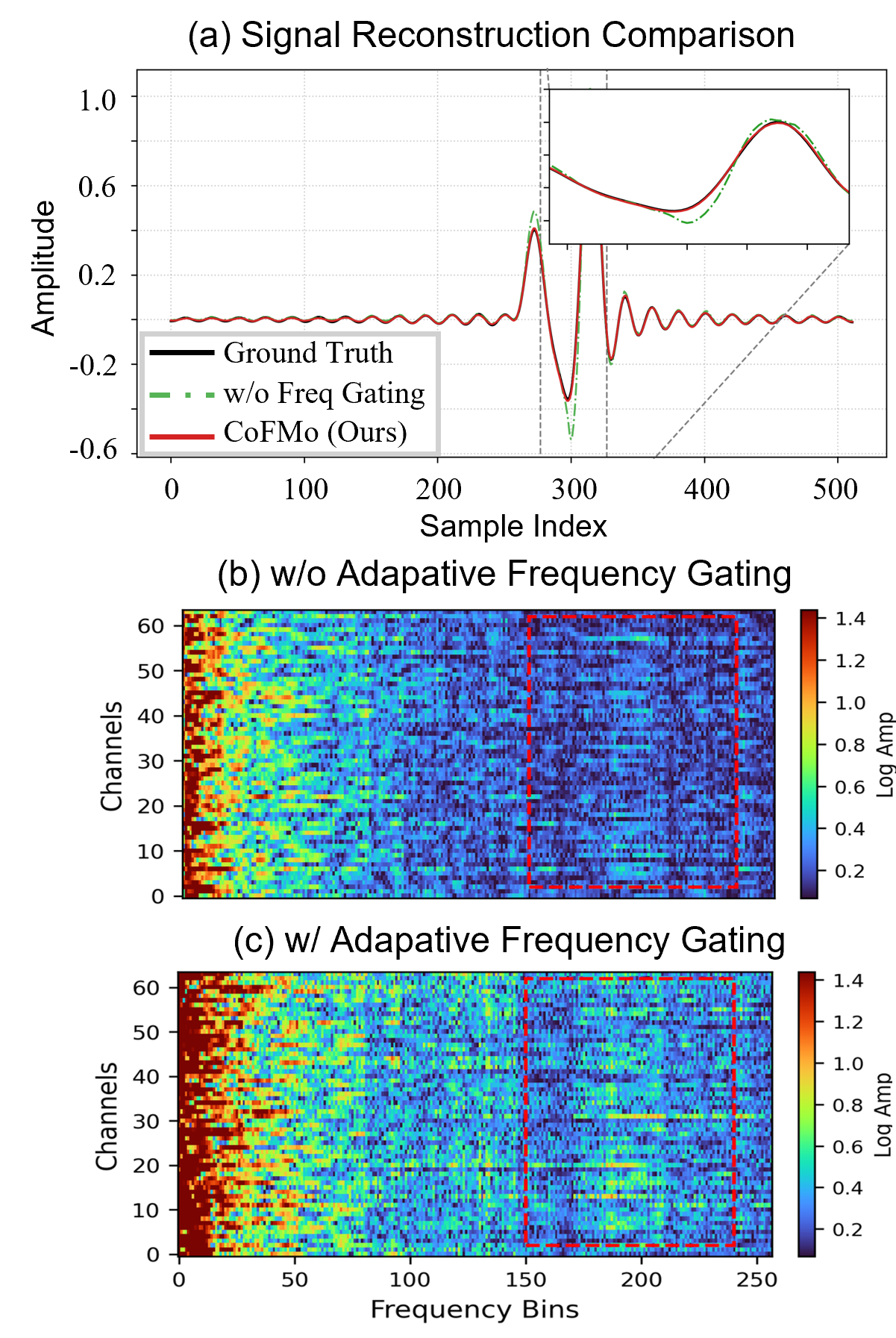}
\caption{\TBD{Visualization of the proposed adaptive frequency gating mechanism. (a) Signal reconstruction ablation. The baseline without frequency gating exhibits severe ringing and overshooting artifacts at abrupt transitions, whereas the proposed CoFMo achieves a smooth, artifact-free reconstruction. (b) and (c) represent the magnitude spectrum heatmaps before and after applying the adaptive gating, respectively. The red highlighted region shows that the proposed adaptive gating mechanism attenuates unstable high-frequency activations and produces a more compact spectral representation.}}
\label{fig2_4}
\end{figure}

Fig.~\ref{fig2_4} illustrates this effect. Without the gate, the reconstruction exhibits a larger overshoot and more pronounced local ringing near the abrupt transition. In contrast, CoFMo follows the ground truth more closely and reduces these reconstruction artifacts. Moreover, the learned gating mechanism selectively suppresses unstable high-frequency activations, leading to a more compact and structured spectral representation.

The spectral layer captures global spectral structure but not explicit
position-wise temporal dependencies. We therefore transpose the output
$f_{\mathrm{out}}$ to the sequence-oriented form
$f_{\mathrm{seq}} \in \mathbb{R}^{B \times L \times C}$ and apply a
standard multi-head self-attention block~\cite{vaswani2017attention} with
$m=4$ heads, consisting of multi-head self-attention (MHSA) and an MLP,
each followed by layer normalization (LN) and a residual connection,
\begin{equation}
\begin{aligned}
    f_{\mathrm{seq}}^{\prime} &= f_{\mathrm{seq}} + \mathrm{LN}\big(\mathrm{MHSA}(f_{\mathrm{seq}})\big), \\
    f_{\mathrm{tsm}} &= f_{\mathrm{seq}}^{\prime} + \mathrm{LN}\big(\mathrm{MLP}(f_{\mathrm{seq}}^{\prime})\big).
\end{aligned}
\end{equation}
The output $f_{\mathrm{tsm}}$ is transposed back to $\mathbb{R}^{B \times C \times L}$
for the downstream decomposition branch.

\TBD{\textbf{Remark.} Since the multi-head self-attention has a complexity of $\mathcal{O}(L^2)$ in the sequence length $L$, a long or continuous streaming waveform can be recovered segment by segment. The waveform is divided into consecutive segments of fixed length $L$, each segment is recovered by CoFMo, and the recovered segments are concatenated. Adjacent segments share a few overlapping samples, which are used to align their fold counts before concatenation. The total cost therefore grows linearly with the length of the waveform, and the memory is bounded by one segment.}

\subsection{Discrete Decomposition Branch}
The contextual features $f_{\mathrm{tsm}}\in\mathbb{R}^{B\times C\times L}$ produced by the temporo-spectral module encode the long-range dependencies required to disambiguate the folding pattern, yet the recovery of $x[n]$ remains contingent on the integer fold-count sequence $k[n]$. A direct regression of $k[n]$ is numerically fragile because any residual deviation $|\hat{k}[n]-k[n]|>\tfrac{1}{2}$ is rounded to a $2\lambda$-level offset after closed-form unwrapping, which propagates through subsequent samples as a cascading cycle slip. Thus, we cast the coarse estimation as a pair of parallel classification tasks and couple them through a physics constraint.

Let $\mathcal{P}_{k}(\cdot)$ and $\mathcal{P}_{\Delta k}(\cdot)$ denote two convolutional predictors, each realized as a Conv1D, batch normalization, a rectified nonlinearity, and a $1\times 1$ channel-wise projection. They map the shared representation $f_{\mathrm{tsm}}$ to two logit tensors as:

\begin{equation}
\mathbf{Z}_{k} \;=\; \mathcal{P}_{k}(f_{\mathrm{tsm}})
               \;\in\; \mathbb{R}^{B\times(2K_{\max}+1)\times L},
\end{equation}
\begin{equation}
\mathbf{Z}_{\Delta k} \;=\; \mathcal{P}_{\Delta k}(f_{\mathrm{tsm}})     \;\in\; \mathbb{R}^{B\times |\mathcal{I}_{\Delta k}| \times L},
\end{equation}
where $K_{\max}$ and $\Delta k_{\max}$ are the absolute and relative fold budgets. The absolute channel axis indexes the candidate integers $i \in \mathcal{I}_k \triangleq \{-K_{\max}, \dots, K_{\max}\}$, while the relative axis indexes $j \in \mathcal{I}_{\Delta k} \triangleq \{-\Delta k_{\max}, \dots, \Delta k_{\max}\}$ with $\vert{}\mathcal{I}_{\Delta k}\vert{} = 2\Delta k_{\max} + 1$, covering the transitions of $k[n]$ between adjacent samples within the target operating range. Next, sample-wise categorical posteriors are obtained via channel-wise softmax:

\begin{equation}
p_{k,i}[n] = \frac{\exp\!\bigl(\mathbf{Z}_{k,i}[n]\bigr)}  {\displaystyle\sum_{i'\in\mathcal{I}_{k}}
                   \exp\!\bigl(\mathbf{Z}_{k,i'}[n]\bigr)},
\end{equation}

\begin{equation}
p_{\Delta k,j}[n] = \frac{\exp\!\bigl(\mathbf{Z}_{\Delta k,j}[n]\bigr)}
                         {\displaystyle\sum_{j'\in\mathcal{I}_{\Delta k}}
                          \exp\!\bigl(\mathbf{Z}_{\Delta k,j'}[n]\bigr)}
\end{equation}

For downstream coarse unwrapping, we retain the maximum-a-posteriori estimates as:

\begin{equation}
\hat{k}[n] \;=\; \operatorname*{arg\,max}_{i\in\mathcal{I}_{k}}\,p_{k,i}[n],
\qquad
\Delta \hat k[n] \;=\; \operatorname*{arg\,max}_{j\in\mathcal{I}_{\Delta k}}\,
                             p_{\Delta k,j}[n],
                             \label{eq22}
\end{equation}
which parametrize a discrete search over $2K_{\max}+1$ integer hypotheses per sample rather than a continuous real-valued one, eliminating sub-integer error accumulation by construction.

\subsection{Coarse Recovery and Refinement}
Following the discrete decomposition, the absolute fold-count estimate $\hat{k}$ is used to invert the modulo operation: the coarse recovery layer performs a closed-form unwrapping based on the physical formulation of the modulo ADC. The initial coarse estimate $x_{\mathrm{coarse}}$ is reconstructed as:
\begin{equation}
    x_{\mathrm{coarse}} = y_{\mathrm{e}} + 2\lambda\hat{k}.
    \label{eq_coarse}
\end{equation}
This transformation restores the signal's fundamental scale and global trajectory. However, $x_{\mathrm{coarse}}$ relies heavily on the classification branch, as any isolated misclassification in $\hat{k}$ induces a severe $2\lambda$ amplitude offset. Thus, subsequent continuous regression is crucial to correct these artifacts and achieve high-fidelity recovery.

Specifically, the coarse estimate $x_{\mathrm{coarse}}$ is concatenated channel-wise with the contextual representation $f_{\mathrm{tsm}}$ produced by the gated temporo-spectral module as:
\begin{equation}
  f_{\mathrm{reg}}
    = \mathrm{Concat}\!\left(f_{\mathrm{tsm}},\;
                              x_{\mathrm{coarse}}\right)
  \label{eq:freg}
\end{equation}

A regression head is then applied to $f_{\mathrm{reg}}$. The final signal is obtained by adding the residual to the coarse estimate:
\begin{equation}
  \hat{x}[n] = x_{\mathrm{coarse}}[n] + \hat{r}[n]
             = y_{\mathrm{e}}[n] + 2\lambda\,\hat{k}[n] + \hat{r}[n]
  \label{eq:xhat}
\end{equation}

This coarse-to-fine paradigm ensures that the discrete decomposition handles the macroscopic wrapping ambiguity, while the continuous regression resolves sub-interval distortions and misclassification offsets, ultimately yielding a more accurate and robust signal reconstruction.

\subsection{Loss Function}

CoFMo is trained with five complementary loss terms, which supervise the fold-count classification, the structural consistency between the two classification heads, and the waveform regression in the sample and difference domains. Their weights are learned by the adaptive scheme described at the end of this subsection.

\textbf{Absolute fold classification loss.}
The primary discrete supervision signal is the standard cross-entropy loss over the $(2K_{\max}+1)$-class logit tensor $\mathbf{Z}_k$:
\begin{equation}
  \mathcal{L}_{\mathrm{cls}}
    = -\frac{1}{BL}\sum_{b=1}^{B}\sum_{n=1}^{L}
        \log p_{k,\,y_k^{(b)}[n]}^{(b)}[n],
  \label{eq:Lcls}
\end{equation}
where $y_k^{(b)}[n] \in \mathcal{I}_k$ is the ground-truth fold-count class index for sample $b$ at position $n$. Minimising Eq.~(\ref{eq:Lcls}) drives the network to assign high posterior mass to the correct fold count, which is the prerequisite for the coarse estimate via Eq.~(\ref{eq_coarse}).

\textbf{Relative fold-change classification loss.}
An auxiliary cross-entropy loss is applied to the $|\mathcal{I}_{\Delta k}|$-class logit tensor $\mathbf{Z}_{\Delta k}$:
\begin{equation}
  \mathcal{L}_{\Delta\mathrm{cls}}
    = -\frac{1}{BL}\sum_{b=1}^{B}\sum_{n=1}^{L}
        \log p_{\Delta k,\,y_{\Delta k}^{(b)}[n]}^{(b)}[n],
  \label{eq:Ldcls}
\end{equation}
where $y^{(b)}_{\Delta k}[n] \in \{0, 1, \dots, |\mathcal{I}_{\Delta k}| - 1\}$ encodes the ground-truth relative fold-change $\Delta k \in \mathcal{I}_{\Delta k}$ via the shift $y_{\Delta k} = \Delta k - \min(\mathcal{I}_{\Delta k})$. This loss provides a differential supervision signal that is particularly informative at fold-transition boundaries, where the absolute fold count changes by non-zero steps and the local gradient of $y[n]$ exhibits a characteristic discontinuity.

\textbf{Structural consistency loss.}
The absolute and relative fold-count heads are coupled because the first-order finite difference of the absolute fold sequence must equal the relative fold-change at every position. To enforce this constraint in a differentiable and soft manner, we define the soft expected absolute fold count $\bar{k}^{(b)}[n]$ and soft expected relative fold-change $\overline{\Delta k}^{(b)}[n]$ as:
\begin{equation}
\begin{aligned}
  \bar{k}^{(b)}[n] &= \sum_{m=-K_{\max}}^{K_{\max}} m\, p_{k,m}^{(b)}[n], \\
  \overline{\Delta k}^{(b)}[n] &= \sum_{j \in \mathcal{I}_{\Delta k}} j\, p_{\Delta k,j}^{(b)}[n],
\end{aligned}
\label{eq:softexpect}
\end{equation}
where $p_{k,m}^{(b)}[n]$ is the softmax probability assigned to the fold count $m \in \{-K_{\max},\ldots,K_{\max}\}$ for sample $b$ at position $n$, and $p_{\Delta k,j}^{(b)}[n]$ is the corresponding probability for relative fold-change $j$. Unlike the integer MAP estimates in Eq.~\eqref{eq22}, which are used for unwrapping, these real-valued expectations are used only in the consistency loss. We then penalise their mutual inconsistency:
\begin{equation}
  \mathcal{L}_{\mathrm{consist}}
    = \frac{1}{B(L-1)}\sum_{b=1}^{B}\sum_{n=2}^{L}
        \Bigl(
          \bar{k}^{(b)}[n] - \bar{k}^{(b)}[n-1]
          - \overline{\Delta k}^{(b)}[n]
        \Bigr)^{2}.
  \label{eq:Lconsist}
\end{equation}

This loss avoids contradictory predictions that individually satisfy Eq. (\ref{eq:Lcls}) and Eq. (\ref{eq:Ldcls}) but violate the constraint imposed by the modulo operator.

\textbf{Signal regression loss.}
Let $x_{\mathrm{true}}^{(b)}[n]$ denote the ground-truth signal. The pointwise fidelity objective is the mean squared error between the final reconstruction and the ground truth:
\begin{equation}
  \mathcal{L}_{\mathrm{reg}}
    = \frac{1}{BL}\sum_{b=1}^{B}\sum_{n=1}^{L}
        \bigl(\hat{x}^{(b)}[n] - x_{\mathrm{true}}^{(b)}[n]\bigr)^{2}.
  \label{eq:Lreg}
\end{equation}

\textbf{Difference regression loss.}
Eq. (\ref{eq:Lreg}) is insensitive to slope errors and inter-sample discontinuities, which are the dominant artefacts at fold boundaries where $\hat{k}[n]$ transitions between adjacent integer values. To penalise such artefacts, we introduce a difference-domain loss:
\begin{equation}
\begin{aligned}
  \mathcal{L}_{\mathrm{diff}}
    &= \frac{1}{B(L-1)}\sum_{b=1}^{B}\sum_{n=2}^{L}
        \Bigl[
          \bigl(\hat{x}^{(b)}[n]-\hat{x}^{(b)}[n-1]\bigr) \\
  &\qquad\qquad
          - \bigl(x_{\mathrm{true}}^{(b)}[n]-x_{\mathrm{true}}^{(b)}[n-1]\bigr)
        \Bigr]^{2}
\end{aligned}
  \label{eq:Ldiff}
\end{equation}

\begin{figure*}[t]
\centering
\scalefig{0.82}{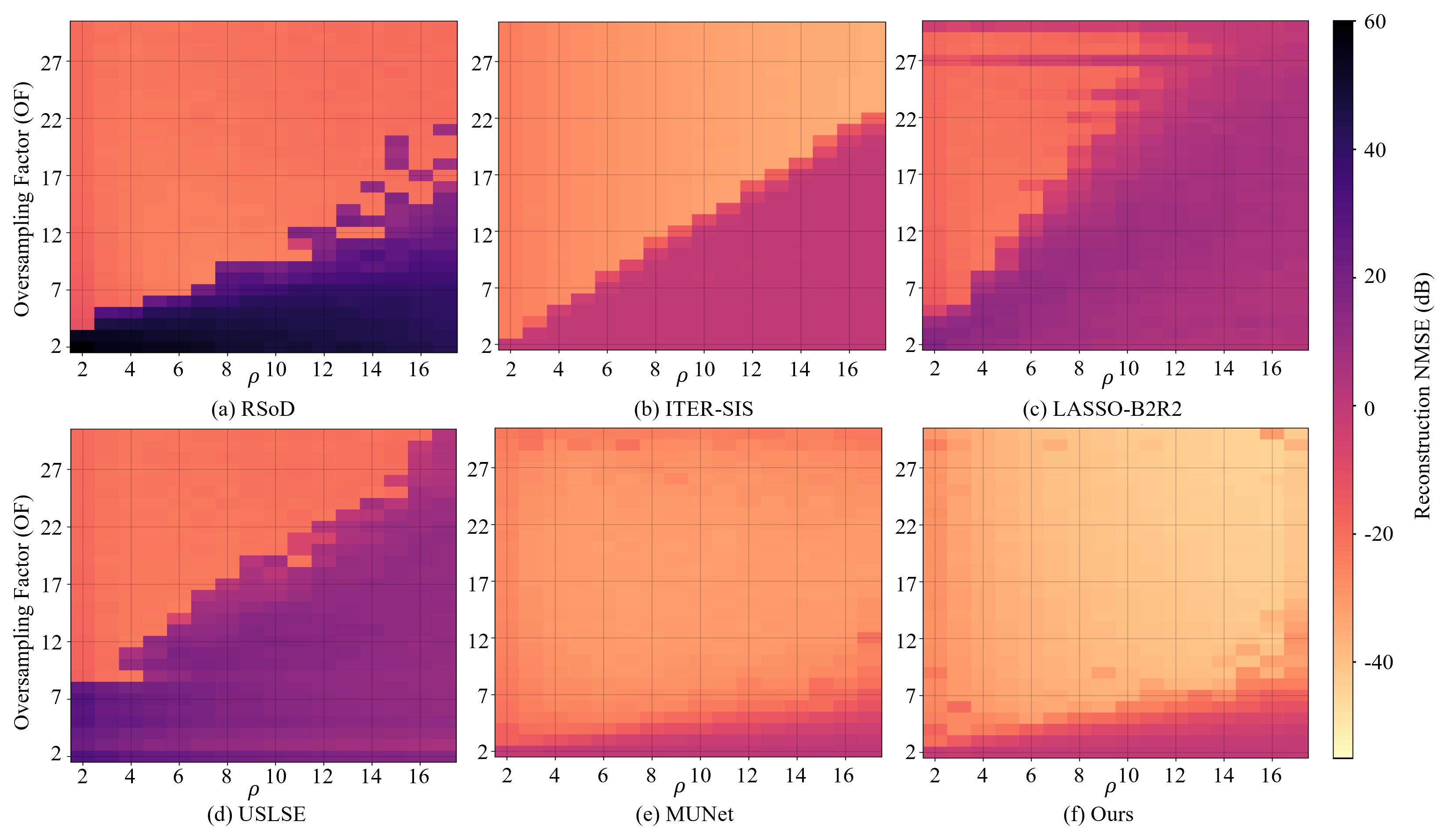}
\caption{\TBD{Phase transition diagrams of the reconstruction NMSE for different modulo unfolding algorithms. The performance is evaluated over the folding ratio $\rho \in [2, 17]$ and the oversampling factor $\mathrm{OF} \in [2, 30]$ at 19~dB SNR with 4-bit quantization, with $100$ test samples averaged per $(\rho, \mathrm{OF})$ pair. The subfigures correspond to (a) RSoD, (b) ITER-SIS, (c) LASSO-B2R2, (d) USLSE, (e) MUNet, and (f) the proposed CoFMo. Lighter regions indicate lower NMSE and highly successful signal recovery.}}
\label{fig3_0}
\end{figure*}

\begin{table*}[t]
\renewcommand{\arraystretch}{1.25}
\centering
\caption{\TBD{Reconstruction NMSE (dB) on eleven representative $(\mathrm{OF},\rho)$ configurations, reported as mean $\pm$ standard deviation. The best result of each row is in \textbf{bold}.}}
\label{tab:representative_nmse}
\scalebox{0.8}{%
\begin{tabular}{c c | c c c c c c}
\hline
\multicolumn{2}{c|}{Configuration} & \multicolumn{6}{c}{NMSE (dB) $\downarrow$} \\
\cline{1-2} \cline{3-8}
OF & $\rho$ & RSoD \cite{b2} & ITER-SIS \cite{guo2023iter} & LASSO-B2R2 \cite{shah2023lasso} & USLSE \cite{zhang2024line} & MUNet \cite{b11} & Ours \\
\hline
2 & 2 & 45.05 $\pm$ 23.73 & \textbf{-0.58 $\pm$ 0.88} & 8.83 $\pm$ 10.20 & 26.63 $\pm$ 3.26 & 5.99 $\pm$ 3.18 & 2.33 $\pm$ 9.52 \\
3 & 6 & 41.23 $\pm$ 16.50 & -0.19 $\pm$ 0.18 & 7.42 $\pm$ 4.89 & 11.94 $\pm$ 3.63 & -2.70 $\pm$ 4.25 & \textbf{-18.54 $\pm$ 4.28} \\
4 & 12 & 38.67 $\pm$ 10.78 & -0.23 $\pm$ 0.25 & 4.21 $\pm$ 3.22 & 8.59 $\pm$ 3.15 & -3.10 $\pm$ 3.85 & \textbf{-22.21 $\pm$ 3.98} \\
5 & 17 & 35.36 $\pm$ 12.98 & -0.20 $\pm$ 0.17 & 3.28 $\pm$ 2.67 & 9.21 $\pm$ 3.89 & -2.94 $\pm$ 3.37 & \textbf{-16.58 $\pm$ 2.29} \\
10 & 5 & -24.07 $\pm$ 1.54 & -27.80 $\pm$ 0.47 & -21.87 $\pm$ 6.86 & -11.15 $\pm$ 17.30 & -15.45 $\pm$ 4.02 & \textbf{-34.87 $\pm$ 1.88} \\
10 & 10 & -25.78 $\pm$ 3.64 & -0.69 $\pm$ 0.79 & 5.05 $\pm$ 9.87 & 4.25 $\pm$ 15.04 & -14.67 $\pm$ 4.11 & \textbf{-39.43 $\pm$ 1.51} \\
15 & 2 & -19.03 $\pm$ 1.25 & -23.80 $\pm$ 0.82 & -18.76 $\pm$ 1.20 & -12.97 $\pm$ 9.23 & -14.23 $\pm$ 4.81 & \textbf{-30.27 $\pm$ 1.79} \\
15 & 10 & -24.56 $\pm$ 3.98 & -32.23 $\pm$ 0.97 & -6.41 $\pm$ 16.60 & -7.97 $\pm$ 17.59 & -14.08 $\pm$ 5.08 & \textbf{-40.69 $\pm$ 1.57} \\
15 & 17 & -20.88 $\pm$ 11.67 & -0.79 $\pm$ 0.87 & 4.33 $\pm$ 4.36 & 6.91 $\pm$ 9.01 & -11.04 $\pm$ 3.81 & \textbf{-34.47 $\pm$ 3.93} \\
20 & 10 & -23.21 $\pm$ 3.81 & -31.93 $\pm$ 0.85 & -16.98 $\pm$ 11.19 & -14.39 $\pm$ 11.51 & -11.59 $\pm$ 3.90 & \textbf{-41.86 $\pm$ 1.62} \\
30 & 17 & -20.39 $\pm$ 2.66 & -35.35 $\pm$ 1.13 & -5.44 $\pm$ 10.23 & -6.41 $\pm$ 10.37 & -8.17 $\pm$ 3.31 & \textbf{-37.91 $\pm$ 2.85} \\
\hline
\end{tabular}}
\end{table*}

\textbf{Adaptive Multi-loss Weighting.} The five losses $\{\mathcal{L}_{\mathrm{cls}},\,\mathcal{L}_{\Delta\mathrm{cls}},\,\mathcal{L}_{\mathrm{consist}},\,\mathcal{L}_{\mathrm{reg}},\, \mathcal{L}_{\mathrm{diff}}\}$ are heterogeneous. Two are cross-entropy classification objectives, two are mean-squared-error regression objectives, and one is a structural consistency regularizer. Their numerical scales and gradient magnitudes differ by orders of magnitude, and their relative importance varies with the signal, since rapidly oscillating signals fold more often and rely more on fold classification, whereas smoother signals rely mainly on waveform regression. Hardware distortions further alter this balance. A fixed-weight combination can be expressed as
\begin{equation}
  \mathcal{L}_{\mathrm{naive}} = \sum_{i=1}^{5} w_i\,\mathcal{L}_i
  \label{eq:Lnaive}
\end{equation}

This equation is suboptimal, since no single choice of $\{w_i\}$ is suitable for all signal realizations and measurement conditions, and an exhaustive grid search over all five weights is computationally prohibitive. To resolve this issue, we parameterize the loss weights by modeling the task-specific homoscedastic uncertainty~\cite{b12} rather than relying on manual tuning. For our regression objectives with observation noise $\sigma_i > 0$, we define the negative log-likelihood as:
\begin{equation}
  -\log p\!\left(y \mid f^{\mathbf{W}}(x)\right)
    \propto \frac{1}{2\sigma_i^2}\,\mathcal{L}_i + \log\sigma_i.
  \label{eq:gauss_nll}
\end{equation}
Similarly, for our classification tasks, applying Boltzmann likelihood yields the analogous expression:
\begin{equation}
  -\log p\!\left(y=c \mid f^{\mathbf{W}}(x),\sigma_i\right)
    \approx \frac{1}{\sigma_i^2}\,\mathcal{L}_i + \log\sigma_i.
  \label{eq:boltz_nll}
\end{equation}
To ensure numerical stability during optimization, we reparameterize each task's noise variance as a learnable scalar $s_i \triangleq \log\sigma_i^2$ initialized to zero, so that the effective weight $\exp(-s_i)=1/\sigma_i^2$ remains strictly positive for all $s_i \in \mathbb{R}$. Substituting this into Eq.~(\ref{eq:gauss_nll}) and Eq.~(\ref{eq:boltz_nll}) and summing over all five tasks yields our unified total loss:
\begin{equation}
  \mathcal{L}_{\mathrm{total}} = \sum_{i=1}^{5} \Bigl[\exp(-s_i)\cdot\mathcal{L}_i + s_i\Bigr], \label{eq:Ltotal}
\end{equation}
where the index-to-loss mapping is defined as $i \in \{1,\ldots,5\} \mapsto \{\mathcal{L}_{\mathrm{cls}},\,\mathcal{L}_{\Delta\mathrm{cls}},\,\mathcal{L}_{\mathrm{consist}},\,\mathcal{L}_{\mathrm{reg}},\,\mathcal{L}_{\mathrm{diff}}\}$. The dynamic weight $\exp(-s_i)$ automatically scales down difficult tasks, while the additive $s_i$ term prevents the degenerate solution $s_i \to +\infty$, and all five scalars are jointly optimized with the network parameters via back-propagation.

\section{Simulation and Measurement Results}
\label{sec:results}
CoFMo is implemented in PyTorch and trained on a single NVIDIA RTX A6000 GPU for 400 epochs with a batch size of 64. We use the Adam optimizer with an initial learning rate of $1\times10^{-3}$, which is halved at epochs 200, 300, and 350. \TBD{For both the simulated and measured datasets, we set $K_{\max}=15$ and $\mathcal{I}_{\Delta k}=\{-12,\dots,12\}$ according to the target operating range, leaving a margin above the largest fold count and fold-count change.}

We evaluate CoFMo using both simulated data and real hardware measurements, and compare it against several state-of-the-art baselines: the analytical RSoD \cite{b2}, ITER-SIS \cite{guo2023iter}, LASSO-B2R2 \cite{shah2023lasso}, and USLSE \cite{zhang2024line}, together with the learning-based MUNet \cite{b11}\TBD{, which is retrained from scratch on the same training data as CoFMo in every experiment.} The reconstruction accuracy is measured by the NMSE in decibels,
\begin{equation}
\mathrm{NMSE}
= 10\log_{{10}}
\frac{\sum_{n=1}^{N} \left|x[n]-\hat{x}[n]\right|^2}
{\sum_{n=1}^{N} \left|x[n]\right|^2},
\label{eq:nmse}
\end{equation}
where $\hat{x}[n]$ denotes the reconstruction result, and $x[n]$ is the clean reference before modulo folding and quantization.

\color{black}

Because NMSE aggregates fold and sub-fold errors into a single measure, we also report two fold-count metrics. The fold count implied by a reconstruction is
\begin{equation}
\hat{k}[n]=\mathrm{round}\!\left(\frac{\hat{x}[n]-y_{\mathrm{e}}[n]}{2\lambda}\right),
\label{eq:khat}
\end{equation}
from which we compute the sample-wise fold-count accuracy $\frac{1}{L}\vert\{n\in\{1,\dots,L\}\mid\hat{k}[n]=k[n]\}\vert$ and the exact-unwrapping rate, defined as the fraction of records in which $\hat{k}[n]=k[n]$ holds for all $n$. Under exact unwrapping, the reconstruction error is bounded by the perturbation $\varepsilon[n]$, whereas any fold error incurs an amplitude bias of at least $2\lambda$.

\color{black}

\begin{figure}[t]
\centering
\def\figZeroScale{1.0} %
\def\figZeroLeftShift{+0.01\linewidth} %
\makebox[\linewidth][l]{\hspace*{\figZeroLeftShift}\includegraphics[width=\figZeroScale\linewidth]{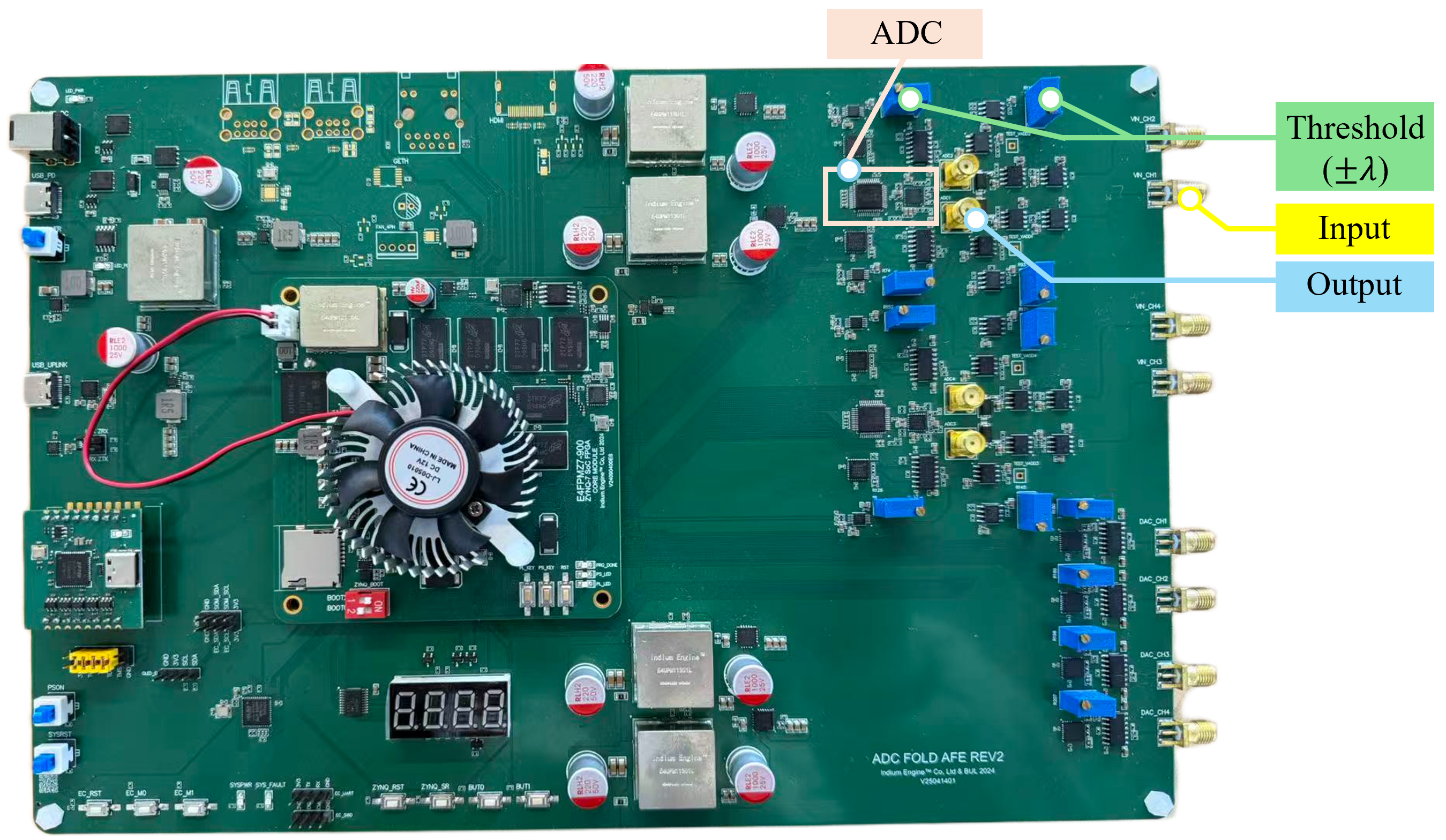}}
\caption{\textcolor{black}{Photograph of our prototype. The labels indicate the on-board ADC, the bipolar threshold generator that sets $\pm\lambda$, and the analog input and modulo output connectors.}}

\label{fig_board}
\end{figure}

\subsection{Simulation Results}

We generate bandlimited signals as random superpositions of shifted sinc bases:
\begin{equation}
x(t)=\sum_{i=1}^{6} a_i \,\mathrm{sinc}(t-i),
\end{equation}
with $a_i \sim \mathcal{U}[-1,1]$ and normalized to $\|x(t)\|_\infty = 1$. Each sequence has $512$ samples and a bandwidth of $0.5$ Hz, giving a Nyquist rate of $f_{\text{Nyq}} = 1$ Hz. \TBD{The main study focuses on finite-energy bandlimited signals. To examine whether the learned fold-count decomposition is restricted to this signal class, additional experiments on time-limited finite-rate-of-innovation (FRI) signals are provided in the supplementary material.}

The modulo samples are quantized to 4 bits and corrupted by additive white Gaussian noise (AWGN) at $\mathrm{SNR}_y = 19$~dB. This setting matches the SNR of about 19~dB measured on the hardware prototype in~\cite{guo2024sub}. We evaluate reconstruction performance across different combinations of the modulo folding parameter $\rho \in \{2, 3, \dots, 17\}$ and the oversampling factor $\text{OF} \in \{2, 3, \dots, 30\}$. We generate a dataset comprising $100000$ independent and identically distributed samples for training. For evaluation, we construct a testing set by generating $100$ independent samples for each unique combination of $\rho$ and OF, yielding a total of $46400$ test samples.

Fig.~\ref{fig3_0} shows the NMSE phase-transition diagrams. CoFMo attains the broadest successful-recovery region. The analytical methods exhibit a diagonal phase boundary and require a proportionally higher OF as $\rho$ grows, which limits them at constrained sampling rates. MUNet improves over them but still covers a smaller region than CoFMo.

\TBD{Table~\ref{tab:representative_nmse} shows the NMSE on eleven representative configurations. CoFMo attains the lowest NMSE in ten of them. For OF from 3 to 5, the NMSE of every baseline stays above $-3.2$~dB, whereas CoFMo reaches $-16.58$ to $-22.21$~dB. Table~\ref{tab:representative_fold} explains this gap. At $(\mathrm{OF},\rho)=(4,12)$, MUNet recovers 95.75\% of the fold counts but no complete record, and each remaining fold error adds an offset of at least $2\lambda$. CoFMo recovers 98.75\% of the fold counts and 42\% of the records exactly. All methods fail at $\mathrm{OF}=2$.
}

\begin{table*}[t]
\renewcommand{\arraystretch}{1.25}
\centering
\caption{\TBD{Exact-unwrapping rate (\%) / sample-wise fold-count accuracy (\%) on the same eleven configurations, obtained from the implied fold counts of \eqref{eq:khat}. Higher is better; the best result of each row is in \textbf{bold}.}}
\label{tab:representative_fold}
\scalebox{0.78}{%
\begin{tabular}{c c | c c c c c c}
\hline
\multicolumn{2}{c|}{Configuration} & \multicolumn{6}{c}{Exact unwrapping (\%) $\uparrow$ / Fold-count accuracy (\%) $\uparrow$} \\
\cline{1-2} \cline{3-8}
OF & $\rho$ & RSoD \cite{b2} & ITER-SIS \cite{guo2023iter} & LASSO-B2R2 \cite{shah2023lasso} & USLSE \cite{zhang2024line} & MUNet \cite{b11} & Ours \\
\hline
2 & 2 & 9 / 13.18 & 8 / 95.72 & 9 / 67.62 & 0 / 19.19 & 0 / \textbf{96.46} & \textbf{13} / 95.77 \\
3 & 6 & 2 / 4.11 & 0 / 86.51 & 0 / 41.35 & 0 / 26.71 & 1 / 97.67 & \textbf{44 / 99.28} \\
4 & 12 & 0 / 0.14 & 0 / 73.72 & 0 / 30.01 & 0 / 17.22 & 0 / 95.75 & \textbf{42 / 98.75} \\
5 & 17 & 2 / 2.16 & 0 / 63.52 & 0 / 25.38 & 0 / 10.76 & 0 / 93.47 & \textbf{13 / 97.60} \\
10 & 5 & \textbf{100 / 100.00} & \textbf{100 / 100.00} & 96 / 96.01 & 61 / 81.66 & 46 / 99.03 & \textbf{100 / 100.00} \\
10 & 10 & 97 / 99.91 & 0 / 75.06 & 8 / 21.87 & 16 / 38.07 & 3 / 96.50 & \textbf{100 / 100.00} \\
15 & 2 & \textbf{100 / 100.00} & \textbf{100 / 100.00} & \textbf{100 / 100.00} & 71 / 97.53 & 81 / 99.60 & \textbf{100 / 100.00} \\
15 & 10 & 91 / 99.65 & \textbf{100 / 100.00} & 41 / 50.07 & 46 / 70.00 & 2 / 94.30 & \textbf{100 / 100.00} \\
15 & 17 & 32 / 92.23 & 0 / 55.43 & 0 / 14.89 & 0 / 25.34 & 0 / 87.62 & \textbf{44 / 99.46} \\
20 & 10 & 79 / 99.19 & \textbf{100 / 100.00} & 58 / 80.83 & 51 / 87.72 & 0 / 88.52 & \textbf{100 / 100.00} \\
30 & 17 & 7 / 88.49 & \textbf{100 / 100.00} & 2 / 36.24 & 3 / 58.24 & 0 / 67.00 & 45 / 99.54 \\
\hline
\end{tabular}}
\end{table*}

\begin{table}[t]
\renewcommand{\arraystretch}{1.2}
\centering
\caption{\TBD{Computational complexity and runtime per $L=512$ sequence. The runtime is given as the range over $(\mathrm{OF},\rho)\in\{(2,2),(15,2),(15,10),(15,17),(30,17)\}$ at a batch size of one, and one multiply-accumulate operation is counted as two FLOPs. The analytical methods have no trainable parameters, and their operation counts depend on the number of iterations.}}
\label{tab:complexity}
\def\complexityTableScale{0.85} %
\resizebox{\complexityTableScale\linewidth}{!}{%
\begin{tabular}{lccc}
\toprule
Method & Params (M) & FLOPs (G) & Runtime (ms) \\
\midrule
RSoD \cite{b2} & -- & -- & 0.09--0.11 \\
LASSO-B2R2 \cite{shah2023lasso} & -- & -- & 68--190 \\
ITER-SIS \cite{guo2023iter} & -- & -- & 1596--2614 \\
USLSE \cite{zhang2024line} & -- & -- & 5935--10091 \\
MUNet \cite{b11} & 2.38 & 2.43 & 2.69--2.73 \\
Ours & 1.02 & 1.10 & 5.96--5.98 \\
\bottomrule
\end{tabular}}
\end{table}

\begin{table*}[!t]
\renewcommand{\arraystretch}{1.3}
\centering
\caption{\TBD{Performance comparison on measured general bandlimited signals without
deliberately injected noise, reported as NMSE, fold-count accuracy and
exact-unwrapping rate. Best results are highlighted in \textbf{bold}.}}
\label{tab:nmse_comparison}
\def\measuredNmseTableScale{0.80} %
\resizebox{\measuredNmseTableScale\textwidth}{!}{%
\begin{tabular}{l | c c | c c c c c c}
\hline
\multirow{2}{*}{Metric} & \multicolumn{2}{c|}{Signal Parameters} & \multicolumn{6}{c}{Recovery method} \\
\cline{2-3} \cline{4-9}
 & $\rho$ & OF & RSoD \cite{b2} & LASSO-B2R2 \cite{shah2023lasso} & ITER-SIS \cite{guo2023iter} & USLSE \cite{zhang2024line} & MUNet \cite{b11} & Ours \\
\hline
\multirow{3}{*}{NMSE (dB) $\downarrow$} & [3.67, 7.08] & [3.36, 3.86] & 16.54 & 0.49 & -0.13 & 9.35 & -16.88 & \textbf{-28.45} \\
 & [8.67, 12.25] & [3.10, 3.51] & 11.09 & -0.14 & -0.05 & 4.53 & -16.20 & \textbf{-29.81} \\
 & [13.08, 17.72] & [4.24, 4.86] & 8.35 & 0.15 & -0.38 & 2.33 & -14.74 & \textbf{-25.96} \\
\cline{1-9}
\multirow{3}{*}{\begin{tabular}[c]{@{}l@{}}Fold-count\\ acc. (\%) $\uparrow$\end{tabular}} & [3.67, 7.08] & [3.36, 3.86] & 12.1 & 77.3 & 71.5 & 22.6 & 85.6 & \textbf{98.8} \\
 & [8.67, 12.25] & [3.10, 3.51] & 6.3 & 22.0 & 29.6 & 20.6 & 83.0 & \textbf{96.7} \\
 & [13.08, 17.72] & [4.24, 4.86] & 4.5 & 27.0 & 26.8 & 13.1 & 79.8 & \textbf{94.2} \\
\cline{1-9}
\multirow{3}{*}{\begin{tabular}[c]{@{}l@{}}Exact unwrap.\\ (\%) $\uparrow$\end{tabular}} & [3.67, 7.08] & [3.36, 3.86] & 0.0 & 3.3 & 3.3 & 0.0 & 6.7 & \textbf{43.3} \\
 & [8.67, 12.25] & [3.10, 3.51] & 0.0 & 0.0 & 0.0 & 0.0 & 3.3 & \textbf{23.3} \\
 & [13.08, 17.72] & [4.24, 4.86] & 0.0 & 0.0 & 0.0 & 0.0 & 0.0 & \textbf{13.3} \\
\hline
\end{tabular}
}
\end{table*}

\begin{table*}[t]
\renewcommand{\arraystretch}{1.3}
\centering
\caption{\TBD{Performance comparison on measured general bandlimited signals under noisy conditions and 4-bit quantization, reported as NMSE, fold-count accuracy and exact-unwrapping rate. Best results are highlighted in \textbf{bold}.}}
\label{tab:nmse_noisy_quantized}
\def\nmseTableScale{0.83} %
\resizebox{\nmseTableScale\textwidth}{!}{%
\begin{tabular}{l | c | c c | c c c c c c}
\hline
\multirow{2}{*}{Metric} & \multirow{2}{*}{\begin{tabular}[c]{@{}c@{}}$\mathrm{SNR}_{x}$\\ (dB)\end{tabular}} & \multicolumn{2}{c|}{Signal Parameters} & \multicolumn{6}{c}{Recovery method} \\
\cline{3-4} \cline{5-10}
 & & $\rho$ & OF & RSoD \cite{b2} & LASSO-B2R2 \cite{shah2023lasso} & ITER-SIS \cite{guo2023iter} & USLSE \cite{zhang2024line} & MUNet \cite{b11} & Ours \\
\hline
\multirow{6}{*}{NMSE (dB) $\downarrow$} & \multirow{2}{*}{10} & [6.44, 10.75] & [3.63, 4.12] & 11.36 & 0.39 & -0.06 & 4.66 & -13.32 & \textbf{-21.58} \\
 &  & [12.17, 16.62] & [4.95, 5.53] & 12.85 & 0.28 & -0.04 & 3.82 & -11.64 & \textbf{-19.45} \\
\cline{2-10}
 & \multirow{2}{*}{15} & [6.19, 10.56] & [2.77, 3.24] & 15.59 & 0.07 & -0.06 & 5.60 & -13.82 & \textbf{-23.15} \\
 &  & [13.08, 18.52] & [4.02, 4.57] & 16.24 & 0.13 & -0.14 & 4.15 & -12.95 & \textbf{-22.80} \\
\cline{2-10}
 & \multirow{2}{*}{20} & [7.08, 11.64] & [4.14, 4.64] & 14.12 & 0.05 & -0.22 & 4.31 & -18.75 & \textbf{-29.64} \\
 &  & [13.86, 17.38] & [4.27, 4.86] & 7.47 & 0.02 & -0.38 & 2.84 & -17.20 & \textbf{-27.85} \\
\cline{1-10}
\multirow{6}{*}{\begin{tabular}[c]{@{}l@{}}Fold-count\\ acc. (\%) $\uparrow$\end{tabular}} & \multirow{2}{*}{10} & [6.44, 10.75] & [3.63, 4.12] & 6.2 & 39.9 & 47.9 & 23.4 & 71.5 & \textbf{91.2} \\
 &  & [12.17, 16.62] & [4.95, 5.53] & 3.8 & 34.2 & 37.3 & 18.9 & 65.2 & \textbf{87.4} \\
\cline{2-10}
 & \multirow{2}{*}{15} & [6.19, 10.56] & [2.77, 3.24] & 7.4 & 54.2 & 57.3 & 20.8 & 79.4 & \textbf{93.8} \\
 &  & [13.08, 18.52] & [4.02, 4.57] & 4.6 & 48.6 & 49.1 & 17.4 & 76.8 & \textbf{91.5} \\
\cline{2-10}
 & \multirow{2}{*}{20} & [7.08, 11.64] & [4.14, 4.64] & 8.4 & 72.5 & 71.8 & 25.1 & 92.4 & \textbf{98.6} \\
 &  & [13.86, 17.38] & [4.27, 4.86] & 7.9 & 55.9 & 51.3 & 22.4 & 88.7 & \textbf{95.6} \\
\cline{1-10}
\multirow{6}{*}{\begin{tabular}[c]{@{}l@{}}Exact unwrap.\\ (\%) $\uparrow$\end{tabular}} & \multirow{2}{*}{10} & [6.44, 10.75] & [3.63, 4.12] & 0.0 & 0.0 & 0.0 & 0.0 & 3.3 & \textbf{23.3} \\
 &  & [12.17, 16.62] & [4.95, 5.53] & 0.0 & 0.0 & 0.0 & 0.0 & 0.0 & \textbf{3.3} \\
\cline{2-10}
 & \multirow{2}{*}{15} & [6.19, 10.56] & [2.77, 3.24] & 0.0 & 0.0 & 0.0 & 0.0 & 6.7 & \textbf{30.0} \\
 &  & [13.08, 18.52] & [4.02, 4.57] & 0.0 & 0.0 & 0.0 & 0.0 & 0.0 & \textbf{10.0} \\
\cline{2-10}
 & \multirow{2}{*}{20} & [7.08, 11.64] & [4.14, 4.64] & 0.0 & 3.3 & 3.3 & 0.0 & 16.7 & \textbf{56.7} \\
 &  & [13.86, 17.38] & [4.27, 4.86] & 0.0 & 0.0 & 0.0 & 0.0 & 6.7 & \textbf{23.3} \\
\hline
\end{tabular}
}
\end{table*}

\begin{figure}[t]
\centering
\scalefig{1.0}{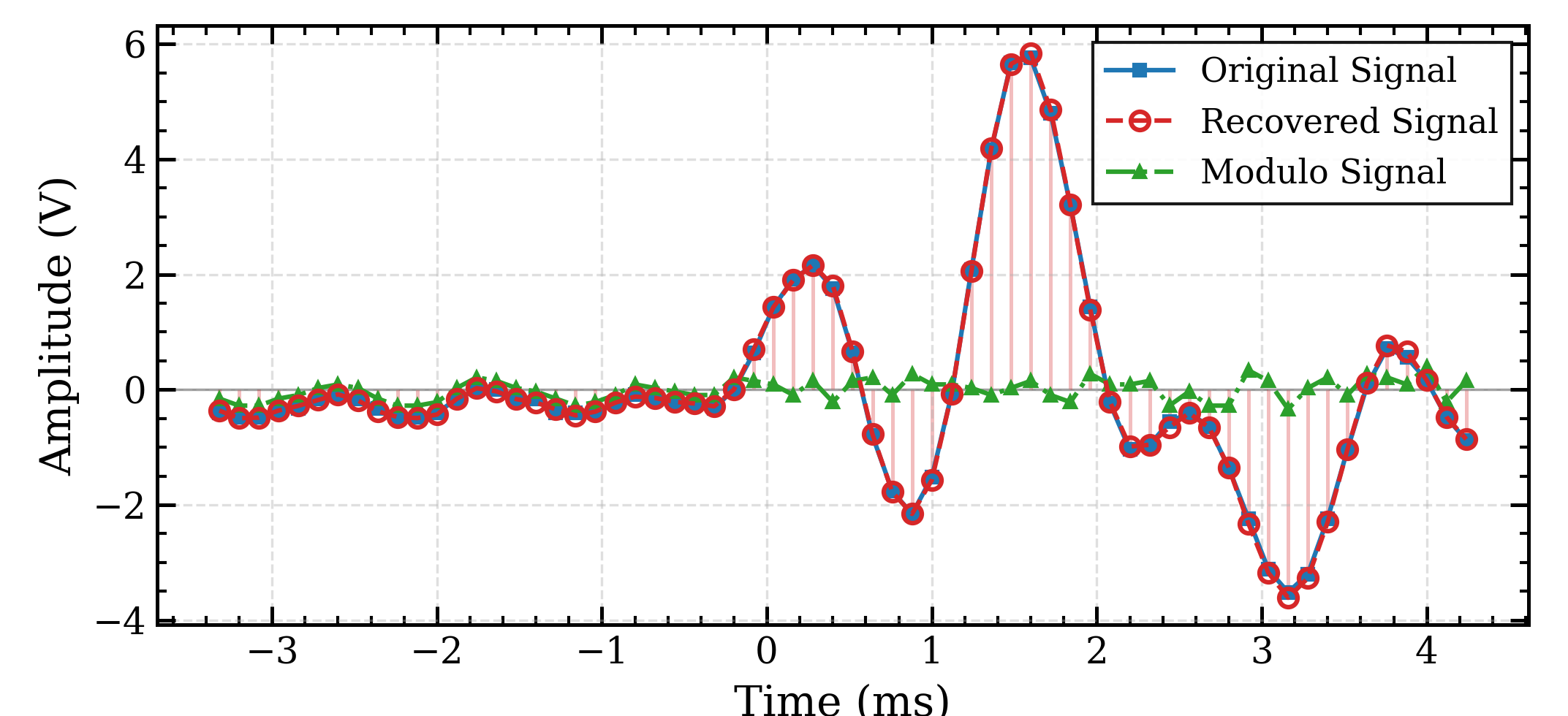}
\caption{Measured waveform recovered by CoFMo with injected noise and 4-bit quantization. The green, blue, and red curves denote the modulo samples, the unfolded reference, and the recovered signal, respectively.}
\label{fig: truedata}
\end{figure}

\color{black}
\subsection{Computational Complexity}
\label{sec:complexity}

Table~\ref{tab:complexity} compares the computational cost of all methods on five of the simulated configurations in Table~\ref{tab:representative_nmse}. CoFMo requires 1.02~M trainable parameters and 1.10~GFLOPs per 512-sample sequence, less than half the parameter count and arithmetic cost of MUNet. Its measured latency is approximately 6~ms per sequence on an NVIDIA RTX A6000 GPU and remains nearly unchanged across the tested folding ratios and oversampling factors, because the recovery uses a fixed forward pass without iterative optimization. The processing time per record is therefore independent of the folding severity and can be budgeted in advance. MUNet has a lower measured latency of about 2.7~ms, whereas the iterative analytical methods LASSO-B2R2, ITER-SIS, and USLSE require from about 68~ms to more than 10~s per sequence. RSoD is the fastest method but fails for OF from 2 to 5 in Table~\ref{tab:representative_nmse}.
\color{black}

\subsection{Measurement Results}

\textcolor{black}{We validate CoFMo on the modulo ADC prototype in Fig.~\ref{fig_board}, which follows the FPGA-calibrated architecture in our previous work~\cite{HardwareFPGA}. Its on-board ADC has enabled real-time recovery for a two-channel modulo ADC with different thresholds~\cite{li2025live, li2026onboard}. In contrast, state-of-the-art single-channel algorithms, including all baselines in this work, are executed offline~\cite{b2, zhang2024line, shah2023lasso, guo2023iter, b11}. For a consistent comparison with them, a single channel is used and the on-board ADC is bypassed, with the folded output recorded by an oscilloscope as in earlier modulo ADC prototypes~\cite{bhandari2021unlimited, guo2023iter, zhu2025modulo, b2}. Real-time on-board recovery is left to future work.}

\begin{table}[t]
\renewcommand{\arraystretch}{1.3}
\centering
\caption{Ablation on the Dual-Stream Initialization (DSI) module: NMSE for General Bandlimited signals under noisy conditions and 4-bit quantization.}
\label{tab:nmse_dsi_ablation}
\def\dsiTableScale{0.80} %
\resizebox{\dsiTableScale\linewidth}{!}{%
\begin{tabular}{c | c c | c c}
\hline
\multirow{2}{*}{\begin{tabular}[c]{@{}c@{}}$\mathrm{SNR}_{x}$\\ (dB)\end{tabular}} & \multicolumn{2}{c|}{Signal Parameters} & \multicolumn{2}{c}{NMSE (dB)} \\
\cline{2-3} \cline{4-5}
& $\rho$ & OF & W/O DSI & W/ DSI \\
\hline
\multirow{2}{*}{10}
& [6.44, 10.75]  & [3.63, 4.12] & $-19.18$ & $-21.58$ \\
& [12.17, 16.62] & [4.95, 5.53] & $-18.71$ & $-19.45$ \\
\hline
\multirow{2}{*}{15}
& [6.19, 10.56]  & [2.77, 3.24] & $-20.39$ & $-23.15$ \\
& [13.08, 18.52] & [4.02, 4.57] & $-21.99$ & $-22.80$ \\
\hline
\multirow{2}{*}{20}
& [7.08, 11.64]  & [4.14, 4.64] & $-28.71$ & $-29.64$ \\
& [13.86, 17.38] & [4.27, 4.86] & $-27.14$ & $-27.85$ \\
\hline
\end{tabular}
}
\end{table}

\begin{table}[t]
\caption{Ablation Study of Architecture Modules and Multi-Loss Weighting Scheme (NMSE in dB)}
\label{tab:ablation_study}
\centering
\def\ablationTableScale{0.80} %
\resizebox{\ablationTableScale\linewidth}{!}{%
\begin{tabular}{lcccc}
\toprule
$\mathrm{SNR}_{x}$ (dB) & w/o Freq Gating & w/o MHSA & w/o AMLW & Full Model \\
\midrule
10 & -19.45 & -18.16 & -18.04 & -20.52 \\
15 & -22.78 & -21.47 & -21.70 & -22.98 \\
20 & -27.77 & -26.94 & -24.99 & -28.75 \\
\bottomrule
\end{tabular}
}
\end{table}

\textcolor{black}{
The test waveforms are synthesized on a host computer and loaded into a Tektronix AFG1022 arbitrary function generator, whose output is split by a coaxial tee between Channel 1 of a Tektronix TDS 1012C EDU oscilloscope, which records the unfolded reference, and the prototype, whose folded output is recorded on Channel 2. All waveforms are random linear combinations of shifted sinc kernels with effective bandwidths from 0.9 to 33 kHz and amplitudes scaled to sweep the folding ratio. Two datasets are acquired: raw measurements, which reflect only the intrinsic nonidealities of the prototype such as loop latency, comparator jitter, and threshold drift, and noisy measurements, for which white Gaussian noise at $\mathrm{SNR}_x = 10, 15,$ and $20\text{ dB}$ is added to the synthesized waveforms before lowpass filtering restores their bandlimited structure. The modulo samples are requantized to 4 bits over $[-\lambda, \lambda)$. Each dataset has 600 training records, and the two test sets have 90 and 180 separately acquired records that follow the training distributions of $\rho$, bandwidth, and $\mathrm{SNR}_x$. Details are given in Section~S1 of the supplementary material.}

\color{black}
\textbf{Experiment \#1.} Table~\ref{tab:nmse_comparison} reports the results on the raw measurements, grouped into three bins with $\rho$ from 3.67 to 17.72 and OF from 3.10 to 4.86. The analytical baselines stay above $-0.4$~dB and fail to resolve the fold ambiguity in such short records at a low OF. CoFMo attains the lowest NMSE in all three bins, from $-25.96$ to $-29.81$~dB, and improves on MUNet by 11.22 to 13.61~dB. It also attains the highest fold-count accuracy and exact-unwrapping rate in every bin. The absolute exact-unwrapping rates are low because a single wrong fold count invalidates the whole record.
\color{black}

\color{black}
\textbf{Experiment \#2.} Table~\ref{tab:nmse_noisy_quantized} reports the results on the second dataset, with $\rho$ from 6.19 to 18.52 and OF from 2.77 to 5.53. The analytical baselines remain near or above 0~dB at all noise levels. CoFMo attains the lowest NMSE in all six groups, from $-19.45$ to $-29.64$~dB, and improves on MUNet by 7.81 to 10.89~dB. In the group with the lowest OF, from 2.77 to 3.24, CoFMo still reaches $-23.15$~dB. It also attains the highest fold-count accuracy, from 87.4\% to 98.6\%, and the highest exact-unwrapping rate in every group. Fig.~\ref{fig: truedata} shows a measured example, in which the recovered waveform follows the unfolded reference despite severe folding, injected noise, and 4-bit requantization.

CoFMo operates on digitized modulo measurements rather than on device-specific internal states. The same architecture can therefore be retrained for other modulo-ADC implementations, provided that the underlying integer-folding structure is preserved.

\color{black}

\subsection{Ablation Study}
We remove or replace each key component of CoFMo to evaluate its contribution. All ablated variants are trained and evaluated under the settings of Experiment~\#2.

\textbf{Effect of Dual-Stream Initialization.} We remove the first-order difference stream $\Delta y[n]$ and retain only the raw modulo samples $y[n]$ as input. As shown in Table~\ref{tab:nmse_dsi_ablation}, the dual-stream initialization (DSI) improves the NMSE by 0.71 to 2.76~dB in all six configurations. At each SNR level, the gain is larger in the first parameter group, which has the lower OF, because the impulse generated by each modulo transition in $\Delta y[n]$ helps the encoder locate fold boundaries that are ambiguous in the raw stream. At a higher OF, the folds are already well resolved in the raw stream, and the gain reduces to 0.71 to 0.81~dB.

\textbf{Effect of the Gated Temporo-Spectral Module.} We remove either the frequency-domain adaptive gating (w/o Freq Gating) or the multi-head self-attention block (w/o MHSA) while retaining the other component. As shown in Table~\ref{tab:ablation_study}, the full model attains the lowest NMSE at every SNR. Removing the gate degrades the NMSE by 1.07, 0.20, and 0.98~dB at $\mathrm{SNR}_{x}=10$, 15, and 20~dB, respectively, whereas removing MHSA causes larger degradations of 2.36, 1.51, and 1.81~dB. MHSA therefore makes the larger contribution, which is consistent with the need to model long-range dependencies in the fold-count sequence, while the gate provides a complementary gain by attenuating noise-contaminated spectral responses.

\textbf{Effect of Adaptive Multi-Loss Weighting.} Replacing the learnable weights with the uniform weights $w_i=1$ in~\eqref{eq:Lnaive}, denoted as w/o AMLW in Table~\ref{tab:ablation_study}, degrades the NMSE by 2.48, 1.28, and 3.76~dB at $\mathrm{SNR}_{x}=10$, 15, and 20~dB, respectively. The uncertainty-driven weighting in~\eqref{eq:Ltotal} therefore balances the heterogeneous losses more effectively than a uniform weighting.

\color{black}
\subsection{Parameter Selection and Sensitivity Analysis}
\label{sec:sensitivity}
This subsection explains how the fold budgets are selected and evaluates the sensitivity of CoFMo to the fold budgets, the random initialization, and the modulo threshold.

\begin{table*}[t]
\renewcommand{\arraystretch}{1.2}
\centering
\caption{\TBD{NMSE (dB) of CoFMo on the configurations of Tables~\ref{tab:representative_nmse} and~\ref{tab:representative_fold} with one parameter changed. $\lfloor(\rho+1)/2\rfloor$ is the largest possible fold count. Each $K_{\max}$ and $\Delta k_{\max}$ setting is a separate training with the data and budget of the selected model. Seeds~1 and~2 retrain the selected configuration. The threshold error $\lambda\rightarrow\lambda(1\pm\varepsilon)$ is applied at test time, and the worse sign is reported. $^{\ast}$Model of Tables~\ref{tab:representative_nmse} and~\ref{tab:representative_fold}.}}
\label{tab:sensitivity}
\def\sensTableScale{0.85} %
\resizebox{\sensTableScale\textwidth}{!}{%
\begin{tabular}{c c c | c c c | c c c | c c c | c c c c}
\hline
\multicolumn{3}{c|}{Configuration} & \multicolumn{3}{c|}{Selected: $K_{\max}{=}15$, $\Delta k_{\max}{=}12$} & \multicolumn{3}{c|}{$K_{\max}$ (with $\Delta k_{\max}{=}12$)} & \multicolumn{3}{c|}{$\Delta k_{\max}$ (with $K_{\max}{=}15$)} & \multicolumn{4}{c}{Threshold error $|\varepsilon|$} \\
\hline
OF & $\rho$ & $\lfloor\frac{\rho+1}{2}\rfloor$ & seed~0$^{\ast}$ & seed~1 & seed~2 & 4 & 8 & 25 & 4 & 8 & 15 & 1\% & 2\% & 5\% & 10\% \\
\hline
2 & 2 & 1 & 2.33 & 1.73 & 2.66 & 0.56 & $-$9.49 & $-$5.43 & 0.65 & 10.30 & $-$3.27 & 2.37 & 2.41 & 2.55 & 2.84 \\
3 & 6 & 3 & $-$18.54 & $-$18.61 & $-$17.26 & $-$18.33 & $-$18.54 & $-$17.17 & $-$20.13 & $-$18.14 & $-$18.26 & $-$18.21 & $-$17.54 & $-$15.48 & $-$13.04 \\
4 & 12 & 6 & $-$22.21 & $-$20.59 & $-$20.35 & $-$18.37 & $-$20.89 & $-$19.55 & $-$24.56 & $-$19.48 & $-$21.95 & $-$21.26 & $-$19.86 & $-$16.92 & $-$14.15 \\
5 & 17 & 9 & $-$16.58 & $-$18.42 & $-$18.68 & $-$16.31 & $-$17.16 & $-$16.52 & $-$20.03 & $-$17.43 & $-$18.76 & $-$16.02 & $-$15.34 & $-$13.81 & $-$12.05 \\
10 & 5 & 3 & $-$34.87 & $-$35.13 & $-$35.37 & $-$34.78 & $-$34.87 & $-$35.15 & $-$35.22 & $-$35.44 & $-$35.56 & $-$33.73 & $-$31.56 & $-$25.90 & $-$20.32 \\
10 & 10 & 5 & $-$39.43 & $-$39.58 & $-$40.06 & $-$36.88 & $-$39.56 & $-$39.63 & $-$39.78 & $-$39.86 & $-$39.96 & $-$36.67 & $-$32.89 & $-$25.87 & $-$20.01 \\
15 & 2 & 1 & $-$30.27 & $-$30.43 & $-$30.62 & $-$30.28 & $-$30.09 & $-$30.41 & $-$30.56 & $-$31.00 & $-$31.19 & $-$29.60 & $-$28.27 & $-$23.94 & $-$18.91 \\
15 & 10 & 5 & $-$40.69 & $-$40.97 & $-$41.31 & $-$37.24 & $-$40.57 & $-$40.79 & $-$41.15 & $-$41.22 & $-$41.44 & $-$37.28 & $-$33.15 & $-$25.94 & $-$20.04 \\
15 & 17 & 9 & $-$34.47 & $-$40.52 & $-$40.90 & $-$37.19 & $-$40.56 & $-$40.48 & $-$41.21 & $-$40.61 & $-$41.23 & $-$31.87 & $-$29.13 & $-$24.01 & $-$19.14 \\
20 & 10 & 5 & $-$41.86 & $-$42.22 & $-$42.51 & $-$37.21 & $-$41.86 & $-$41.96 & $-$42.59 & $-$42.49 & $-$42.67 & $-$37.81 & $-$33.36 & $-$25.98 & $-$20.04 \\
30 & 17 & 9 & $-$37.91 & $-$43.32 & $-$43.53 & $-$35.67 & $-$43.03 & $-$43.28 & $-$44.08 & $-$43.74 & $-$44.02 & $-$34.39 & $-$31.05 & $-$25.03 & $-$19.62 \\
\hline
\end{tabular}}
\end{table*}

\textbf{Fold Budget Bounds.} Both $K_{\max}$ and $\Delta k_{\max}$ are bounded by the signal model. For a signal with peak amplitude $\Vert{}x\Vert{}_\infty = \rho\lambda$, the fold count $k[n] = \lfloor(x[n]+\lambda)/(2\lambda)\rfloor$ satisfies $\vert{}k[n]\vert{} \le \lfloor(\rho+1)/2\rfloor$. By Bernstein's inequality~\cite{b5, nikol1975approximation}, a $\mathcal{B}$-bandlimited signal satisfies $\vert{}x[n]-x[n-1]\vert{} \le \pi\Vert{}x\Vert{}_\infty/\mathrm{OF}$, which with $\vert{}y[n]-y[n-1]\vert{} < 2\lambda$ gives
\begin{equation}
|\Delta k[n]| < \frac{\pi\rho}{2\,\mathrm{OF}} + 1.
\label{eq:dk_bound}
\end{equation}
These bounds give $|k[n]|\le9$ and $|\Delta k[n]|\le14$ for the simulated data with $\rho\le17$ and $\mathrm{OF}\ge2$, and $|k[n]|\le9$ and $|\Delta k[n]|\le11$ for the measured data with $\rho\le18.52$ and $\mathrm{OF}\ge2.77$. The largest values in the $10^{5}$ simulated training sequences are $|k|=8$ and $|\Delta k|=10$. We therefore set $K_{\max}=15$ and $\Delta k_{\max}=12$, which cover all observed fold states with a margin. For a larger dynamic range, $K_{\max}$ is recomputed with the same bound from the largest folding ratio allowed by the input range of the modulo ADC, and the network is retrained.

\textbf{Sensitivity to Fold Budgets and Random Seeds.} Table~\ref{tab:sensitivity} reports the NMSE of CoFMo when one parameter is changed. Once the fold budgets cover the fold range of the data, their exact values have little effect. For $\mathrm{OF}\ge10$, the NMSE with $K_{\max}=8$ or 25 and with $\Delta k_{\max}=4$, 8, or 15 stays within 0.57~dB of the seed range, whereas the NMSE with $K_{\max}=4$, which is below the largest fold count at $\rho\ge10$, is 2.7 to 7.6~dB worse than with $K_{\max}=8$ and 25 at these $\rho$. The small effect of $\Delta k_{\max}$ agrees with the auxiliary role of the relative head in~\eqref{eq:Lconsist}. The three seeds differ by at most 0.66~dB for $\mathrm{OF}\ge10$ and $\rho\le10$. Larger differences occur only at $\rho=17$, where the reported model with seed~0 fails to unwrap a few records, and at $\mathrm{OF}\le5$, where modulo inversion near the Nyquist rate is severely ill-posed.

\textbf{Robustness to Modulo Threshold Mismatch.}
The threshold of the prototype is calibrated as in~\cite{HardwareFPGA} but still drifts with temperature and circuit conditions, from about 0.475 to 0.492~V around the nominal 0.48~V. All hardware results in Tables~\ref{tab:nmse_comparison} and~\ref{tab:nmse_noisy_quantized} use this fixed nominal value without per-record calibration. To quantify the effect of a calibration error, Table~\ref{tab:sensitivity} replaces $\lambda$ by $\lambda(1 \pm \varepsilon)$ in the coarse recovery at test time. The predicted fold counts do not change, and the mismatch adds a bias of $2\lambda\varepsilon\hat{k}[n]$. The NMSE is then approximately $10\log_{10}(10^{\mathrm{NMSE}_0/10} + \varepsilon^2)$, where $\mathrm{NMSE}_0$ is the NMSE without mismatch. A 1\% error costs up to 4 dB in configurations near $-40\text{ dB}$ and less than 1 dB for $\mathrm{OF} \le 5$.

\color{black}

\section{Conclusion}
\label{sec:conclusion}
\textcolor{black}{In this work, we proposed CoFMo, a coarse-to-fine learning framework for modulo signal recovery that disentangles discrete and continuous unwrapping components. Instead of directly regressing the waveform, CoFMo classifies integer fold counts and regresses the sub-fold residual, preventing isolated misclassifications from cascading into $2\lambda$ amplitude offsets. Moreover, a gated temporo-spectral module couples an adaptive frequency-domain design with multi-head self-attention to capture long-range temporal context for reliable fold tracking. Five heterogeneous objectives across classification, structural consistency, and regression are balanced via uncertainty-driven adaptive weighting without manual tuning. Extensive simulations show that CoFMo achieves the broadest recovery region and lowest NMSE among analytical and learning-based baselines. Furthermore, to the best of our knowledge, it is the first learning-based framework to achieve high-fidelity recovery on physical modulo ADC measurements.}

\textcolor{black}{In the present system, the prototype folds the input in real time, and the folded output is recorded by an oscilloscope and recovered offline on a host computer, as in the evaluation of state-of-the-art single-channel methods. CoFMo is positioned as a recovery back end for modulo acquisition systems. Its inference is a fixed forward pass whose cost does not depend on the folding ratio, which is favorable for hardware implementation. Future work will implement CoFMo on the FPGA of the prototype for real-time on-board recovery and extend it to multi-channel and distributed modulo sensing arrays.}

\newcommand{\BIBdecl}{\setlength{\itemsep}{0.25 em}}
\bibliographystyle{IEEEtran}
\itemsep 0pt
\parskip=.1pt
\bibliography{IEEEabrv,refs}

\end{document}